\documentclass[final]{jfp-epi}

\usepackage[T1]{fontenc} 
\usepackage{graphicx}

\usepackage{wrapfig}

\usepackage{listings,lstautogobble}

\usepackage{array}
\newcommand{\PreserveBackslash}[1]{\let\temp=\\#1\let\\=\temp}
\newcolumntype{C}[1]{>{\PreserveBackslash\centering}p{#1}}

\usepackage{tikz}
\usetikzlibrary{patterns}
\usetikzlibrary{arrows.meta, positioning, fit, decorations.pathreplacing, calc}
\usepackage{pgfplots}
\pgfplotsset{width=7cm,compat=1.16}

\usepackage{todonotes}
\usepackage{enumitem}

\usepackage{pifont}

\usepackage{caption}
\usepackage{subcaption}

\newcommand\pbt{PBT}
\newcommand{\namedsystem}[1]{#1}
\newcommand{\codefont}[1]{#1}
\newcommand\qc{\namedsystem{QuickCheck}}

\newcommand\has{\namedsystem{Haskell}}
\newcommand\rocq{\namedsystem{Rocq}}
\newcommand\name{\textsc{Etna}}
\newcommand{\FSUB}{\namedsystem{F$_{<:}$}}

\newcommand{\sectionref}[1]{\S\ref{#1}}
\newcommand{\ourparagraph}[1]{\smallskip {\it #1.}}
\newcommand{\result}[2]{{\em #2}}

\usepackage[group-separator={,}, group-minimum-digits={3}]{siunitx}

\newcommand{\bargraphwidth}{0.45\textwidth}

\usepackage{xcolor}
\usepackage{xspace}
\usepackage{relsize}
\newcommand{\explaincolor}[2]{{\raisebox{-.3ex}{\color{#1}\rule{1.2em}{.8em}}}~= #2}
\newcommand{\explaincolorhatched}[2]{{\raisebox{-.3ex}{\begin{tikzpicture}\fill[#1] (0,0) rectangle (1.2em,.8em); \fill[pattern=north east lines, pattern color=white] (0,0) rectangle (1.2em,.8em);\end{tikzpicture}}}~= #2}

\usepackage{fancyvrb}

\definecolor{dkblue}{rgb}{0,0.1,0.7}
\definecolor{dkgreen}{rgb}{0,0.5,0}
\definecolor{dkred}{rgb}{0.7,0,0}
\definecolor{dkpurple}{rgb}{0.7,0,0.4}
\definecolor{olive}{rgb}{0.4, 0.4, 0.0}
\definecolor{teal}{rgb}{0.0,0.5,0.5}
\definecolor{azure}{rgb}{0.0, 0.4, .8}

\definecolor{chart_black}{HTML}{000000}
\definecolor{chart_red}{HTML}{900D0D}
\definecolor{chart_orange}{HTML}{DC5F00}
\definecolor{chart_blue}{HTML}{243763}
\definecolor{chart_green}{HTML}{436E4F}
\definecolor{chart_purple}{HTML}{6D0E56}
\definecolor{chart_pink}{HTML}{D61C4E}

\definecolor{ltblue}{rgb}{0,0.4,0.4}
\definecolor{dkviolet}{rgb}{0.3,0,0.5}

\lstdefinelanguage{HaskellEtna}{
  sensitive=true,
  alsoletter={'},
  morekeywords={
    case,class,data,default,deriving,do,else,if,import,in,infix,infixl,infixr,
    instance,let,module,newtype,of,then,type,where,qualified,as,hiding,forall,
    family,otherwise
  },
  morekeywords=[2]{
    Int,Integer,Float,Double,Char,Bool,Maybe,Either,IO,String,Ordering,Property,
    Monad,Functor,Applicative,Tree
  },
  morekeywords=[3]{True,False,Nothing,Just,Left,Right,LT,EQ,GT, Node, Leaf},
  morekeywords=[4]{insert,isBST,prop_InsertValid},
  morecomment=[l]--,
  morecomment=[s]{\{-}{-\}},
  morestring=[b]",
}

\lstdefinestyle{haskellstyle}{
  language=HaskellEtna,
  mathescape=true,
  basicstyle=\ttfamily\small,
  showstringspaces=false,
  columns=fullflexible,
  keepspaces=true,
  identifierstyle={\ttfamily\color{black}},
  keywordstyle={\ttfamily\color{dkviolet}},
  keywordstyle=[2]{\ttfamily\color{dkblue}},
  keywordstyle=[3]{\ttfamily\color{ltblue}},
  stringstyle={\ttfamily\color{dkred}},
  commentstyle={\ttfamily\upshape\color{dkgreen}},
}
\lstnewenvironment{hask}{\lstset{style=haskellstyle}}{}

\newcommand{\codeinline}[1]{{\ttfamily\small\textcolor{dkblue}{#1}}}
\renewcommand{\codefont}[1]{\codeinline{#1}}

\lstdefinestyle{bashstyle}{
  language=bash,
  basicstyle=\ttfamily\small,
  showstringspaces=false,
  columns=fullflexible,
  keywordstyle={\ttfamily\color{dkviolet}},
  commentstyle={\ttfamily\upshape\color{dkgreen}},
  stringstyle={\ttfamily\color{dkred}},
}
\lstnewenvironment{bashcode}{\lstset{style=bashstyle}}{}

\lstdefinestyle{ocamlstyle}{
  language=[Objective]Caml,
  basicstyle=\ttfamily\small,
  showstringspaces=false,
  columns=fullflexible,
  keywordstyle={\ttfamily\color{dkviolet}},
  commentstyle={\ttfamily\upshape\color{dkgreen}},
  stringstyle={\ttfamily\color{dkred}},
}
\lstnewenvironment{ocamlcode}{\lstset{style=ocamlstyle}}{}

\lstdefinelanguage{Coq}{
  morekeywords={Fixpoint,match,with,end,fun,let,in,Definition,Theorem,Lemma,Qed,Proof,Inductive,forall},
  sensitive=true,
  morecomment=[s]{(*}{*)},
  morestring=[b]",
}

\lstdefinestyle{rocqstyle}{
  language=Coq,
  mathescape=true,
  basicstyle=\ttfamily\small,
  showstringspaces=false,
  columns=fullflexible,
  keywordstyle={\ttfamily\color{dkviolet}},
  commentstyle={\ttfamily\upshape\color{dkgreen}},
  stringstyle={\ttfamily\color{dkred}},
}
\lstnewenvironment{rocqcode}{\lstset{style=rocqstyle}}{}

\newif\iflater\laterfalse

\newcommand\leo[1]{\textcolor{dkgreen}{\textbf{LEO:} #1}}

\newcommand\hg[1]{\textcolor{dkred}{\textbf{HG:} #1}}

\jfpVolume{36}
\jfpArticle{7}
\jfpDOI{10.46298/jfp.17867}
\jfpYear{2026}
\received[Submitted]{July 2025}
\received[accepted]{May 2026}

\begin{document}

\title{Etna: An evaluation platform for property-based testing}

\author[A. Keles]{Alperen Keles}
\orcid{0009-0000-5734-3598}
\affiliation{%
  \institution{University of Maryland}
  \city{College Park, MD}
  \country{USA}
  \authoremail{akeles@umd.edu}
}

\author[J. Shi]{Jessica Shi}
\orcid{0000-0002-1507-1122}
\affiliation{%
  \institution{University of Pennsylvania}
  \city{Philadelphia, PA}
  \country{USA}
  \authoremail{jwshi@seas.upenn.edu}
}

\author[N. Kamath]{Nikhil Kamath}
\orcid{0009-0004-8612-5329}
\affiliation{%
  \institution{University of Maryland}
  \city{College Park, MD}
  \country{USA}
  \authoremail{nkamath1@umd.edu}
}

\author[T. N. Liu]{Tin Nam Liu}
\orcid{0009-0009-0093-806X}
\affiliation{%
  \institution{University of Pennsylvania}
  \city{Philadelphia, PA}
  \country{USA}
  \authoremail{katrinatinnamliu@gmail.com}
}

\author[C. Mert]{Ceren Mert}
\orcid{0009-0002-9365-0661}
\affiliation{%
  \institution{University of Maryland}
  \city{College Park, MD}
  \country{USA}
  \authoremail{cmert@umd.edu}
}

\author[H. Goldstein]{Harrison Goldstein}
\orcid{0000-0001-9631-1169}
\affiliation{%
  \institution{University at Buffalo}
  \city{Buffalo, NY}
  \country{USA}
  \authoremail{hgoldste@buffalo.edu }
}

\author[B. C. Pierce]{Benjamin C. Pierce}
\orcid{0000-0001-7839-1636}
\affiliation{%
  \institution{University of Pennsylvania}
  \city{Philadelphia, PA}
  \country{USA}
  \authoremail{bcpierce@cis.upenn.edu}
}

\author[L. Lampropoulos]{Leonidas Lampropoulos}
\orcid{0000-0003-0269-9815}
\affiliation{%
  \institution{University of Maryland}
  \city{College Park, MD}
  \country{USA}
  \authoremail{leonidas@umd.edu}
}


\begin{abstract}
  Property-based testing is a mainstay of functional programming, boasting a
  rich literature, an enthusiastic user community, and an abundance of
  tools~--- so many, indeed, that new users may have difficulty
  choosing.  Moreover, any given framework may support a variety of
  strategies for generating test inputs; even experienced users may
  wonder which are better in any given situation. Sadly, the PBT
  literature, though long on creativity, is short on rigorous
  comparisons to help answer such questions.

  We present \name, a platform for empirical evaluation and comparison
  of PBT techniques.  \name{} incorporates a
  number of popular PBT frameworks and testing workloads from the
  literature, and its extensible architecture makes adding new ones easy,
  while handling the technical drudgery of performance measurement.

  To illustrate its benefits, we use
  \name{} to carry out several experiments with popular PBT approaches
  in Rocq, Haskell, OCaml, Racket, and Rust, allowing users to more clearly understand
  best practices and tradeoffs.
\end{abstract}

\maketitle

\section{Introduction}\label{sec:intro}

Haskell's QuickCheck library popularized {\em property-based testing}
(\pbt), which lets users test executable specifications of
their programs by checking them on a large number of
inputs. In fact, QuickCheck made \pbt{} so popular that Claessen and
Hughes's seminal paper~[\citeyear{ClaessenH00}] is the most cited
ICFP paper of all time by a factor of two, according to the ACM Digital Library.
PBT tools can now be found in languages from
OCaml~\citep{QCheckOCaml,Crowbar} and
Scala~\citep{ScalaCheck} to Erlang~\citep{Arts2008,PapadakisS11} and
Python~\citep{Hypothesis}, not to mention proof assistants like
Rocq~\citep{Pierce:SF4}, Agda~\citep{Lindblad07}, and
Isabelle~\citep{Bulwahn12}.

Many aspects of \pbt{} impact its effectiveness, from the properties
themselves~\citep{Hughes2019HowTS} to counterexample
minimization~\citep{HypothesisShrinking}, but arguably the most crucial
one is the algorithm for generating test inputs. Papers citing
QuickCheck often retain its distinctive style of {\em random} test-case
generation, but many other options have been explored. In particular,
{\em enumerative} PBT has also become a staple in the functional
programming community~\citep{RuncimanNL08, RudyMatelaThesis}, and tools
for {\em feedback-based} PBT are gaining ground~\citep{FuzzChick, Crowbar,
  TargetedPBT}. Each of these approaches comes with benefits and
tradeoffs, and choosing one over another can make a big difference on
testing effectiveness.

Even after selecting a generation style~--- say, random
PBT~--- one may be left with quite a few options of {\em framework}, each with
its own unique style. In Haskell, for example, both QuickCheck and
Hedgehog~\citep{Hedgehog} are quite popular.
And even after selecting a framework~--- say, QuickCheck~--- there are
yet more options for choosing a specific generation strategy.
Tools like \namedsystem{generic-random}~\citep{genericRandom} and
\namedsystem{DraGEN}~\citep{MistaR19A} can derive QuickCheck generators
from type information,
offering a quick and accessible entrypoint to \pbt, but their
effectiveness suffers when inputs need to satisfy more complex
semantic constraints.
Alternatively, one can write a {\em bespoke} generator that is
``correct by construction,'' producing only {\em valid} test
inputs. However, such bespoke generators can sometimes become quite
sophisticated~\citep{PalkaAST11, MidtgaardNPFH17,TestingNIjfp}.
And there are other options: for example, QuickChick, Rocq's
variant of QuickCheck, can derive specialized
generators for free from specifications expressed as
inductive relations~\citep{ComputingCorrectly}.
Nuances of the properties under test may make strategies
more or less preferable, and considerable experience may
be required to make a good choice.

Moreover, even after selecting a particular way of using the
tool~--- say, writing a bespoke generator~--- there are
  {\em yet more} options: a given generator can typically be tuned to
produce different sizes and shapes of data.  For example, QuickCheck
generators can be parameterized both globally by a size parameter and
locally by choices like numeric weights on the arguments to various
combinators.

In the existing literature, there are plenty of performance
evaluations for individual PBT tools, but a dearth of {\em comparisons} across
the various available design
dimensions.  New tools are typically evaluated on
just one or two case studies, often showcasing incomparable
measures of effectiveness.
So how is a PBT user supposed to make sense of all these
options?  How is a tool designer supposed to measure success?
How can we turn PBT from an art to a science?

\newcommand*{\notframework}{platform}
Answering these questions is the goal of this paper.  Our
contributions are:
\begin{itemize}
  \item We present \name, an extensible
          {\notframework} for
        evaluating and comparing generation techniques for
        PBT, with generic
        support for measuring performance and presenting results including
        a novel visualization in the form of bucket charts.
        (\sectionref{sec:tool}). \name{} is publicly available at
        \url{https://github.com/alpaylan/etna-cli}.
  \item We populate \name{} with six testing {\em workloads} from the
        literature presenting a range of bug-finding challenges, with
        PBT {\em frameworks} in Haskell, Rocq, OCaml, Racket, and Rust, and with
        various {\em strategies} for using each framework (\sectionref{sec:populating}).
  \item We report on our experiences using \name{} to make
        observations about
        PBT performance. Some of these observations lend
        empirical weight to commonly
        held beliefs, while others suggest improvements to
        existing processes and tools (\sectionref{sec:haskell},
        \sectionref{sec:rocq}, and \sectionref{sec:ocaml}).
  \item We extend \name{} with support for cross-language
        experimentation with popular PBT frameworks in Haskell, Rocq, OCaml,
        Racket, and Rust (\sectionref{sec:cross}), enabling, for the first
        time, precise comparisons of generator efficiency and effectiveness
        across languages.
        %
\end{itemize}

This paper is an extended version of the \citep{etna-icfp23}
experience report published in ICFP 2023. We have since continued our
work on \name, adding support for 3 new languages (OCaml, Racket,
Rust) as well as capabilities for cross-language comparison, with
respective experimentations in \sectionref{sec:ocaml} and \sectionref{sec:cross}. We
have also substantially changed the usage and architecture of \name{},
detailed in \sectionref{sec:tool}. We discuss related and future work in
\sectionref{sec:relwork}.
Since its release, \name{} has already been used to assess the
positive effects of staging and faster randomness in property-based
testing~\cite{richey2025failfasterstagingfast}, while individual
workloads have also been used as case
studies~\cite{QuickerCheck,CombinatorialEnum}.


\section{Platform design}\label{sec:tool}

%

The central purpose of \name{} is to give researchers, library
authors, and expert PBT users an extensible platform for
experimenting with their testing strategies. In this
section, we outline our design principles and the rationale behind
them. Then we describe the \name{} architecture and finish
with a discussion and depiction of communicating with \name{} as a
user.
Before all that, however, we'll begin by providing background on the
generation strategies themselves.

\subsection{Background: Property-based testing and generation strategies}

A key difference between approaches to PBT is how each deals with {\em
    preconditions}.
Consider {\em binary search trees}, where each node value
is greater than everything to its left and less
than everything to its right. In Haskell syntax:
\begin{hask}
  data Tree k v = Leaf | Node (Tree k v) k v (Tree k v)
  isBST  :: Tree k v -> Bool
  insert :: k -> v -> Tree k v -> Tree k v
\end{hask}
What properties should we expect to hold for operations on BSTs such
as \verb|isBST| and \verb|insert|?
\citeauthor{Hughes2019HowTS} thoroughly answers this question
in his guide to writing properties of pure
functions~[\citeyear{Hughes2019HowTS}].  For instance, one desirable property
is that if we insert a key into a valid BST, then it
should remain a valid BST:

\begin{hask}
  prop_InsertValid :: Tree Int () -> Int -> Property
  prop_InsertValid t x = isBST t ==> isBST (insert x () t)
\end{hask}
Here \codeinline{==>} encodes a {\em
    precondition}. That is, the \codeinline{insert} function is only exercised when the
binary tree \codeinline{t} satisfies the \codeinline{isBST} predicate; otherwise,
the property is vacuously true.

There are many ways to
generate data for properties like this.
A simple approach is to straightforwardly follow the structure of the
types to generate arbitrary trees
and filter out the ones that are not BSTs.
While simplistic, this approach works well in some circumstances.  In
fact, for the BST example, such {\em type-driven} approaches can find
all bugs introduced in~\citeauthor{Hughes2019HowTS}'s guide
to writing properties of pure functions~[\citeyear{Hughes2019HowTS}].
But this generate-and-filter approach breaks down with
``sparse''
preconditions that are harder to satisfy randomly; for instance, valid red-black trees are harder to
generate at random than valid BSTs, so type-driven strategies
work less well (see~\sectionref{sec:haskell}
and~\sectionref{sec:rocq}). For yet sparser preconditions,
such as C programs with no undefined
behaviors~\citep{YangCER11}, such an approach is hopeless.
On the other end of the spectrum, users can write {\em
    bespoke generators}: programs that are manually tailored to produce the
desired distribution. Such programs can be extremely effective in finding bugs when the
inputs satisfy the precondition by construction, but they can also be extremely difficult to write.
A well-crafted such generator can in fact be a significant research result: such is
the case for many well-typed term generators in the last
decade~\citep{PalkaAST11, MidtgaardNPFH17,BlockchainTesting,NotUseless}.
Naturally, there are also approaches in the middle. For instance, some
use the structure of the precondition to produce valid data
directly~\citep{Bulwahn12smartgen, ClaessenFLOPS14,FetscherCPHF15,
  Luck, GeneratingGoodGenerators}, while others leverage feedback to
guide generation towards valid or otherwise interesting
inputs~\citep{TargetedPBT, AutomaticTargetedPBT, FuzzChick}.

\subsection{Design principles}

\name{} is designed to help
researchers and framework developers quickly
experiment with different options for PBT data generation. During \name 's development, we focused on a
few key design principles, centered around usefulness, extensibility,
and maintainability.

\subsubsection{Evaluate for the ground truth, not for proxy metrics.}

How do we measure the effectiveness of a generator? The software
testing literature offers two main answers: {\em code coverage} and
  {\em mutation testing}. Code coverage is popular, but problematic: higher
coverage does not always translate to better bug finding~\citep{GopinathJG14,
  EvaluatingFuzzTesting}.
We instead choose
mutation testing~\citep{JiaH11}, which measures the effectiveness of testing by
artificially injecting
mutations to the system under test and checking if testing is
able to detect them. Mutations in the literature~\citep{Hazimeh:2020:Magma, EvaluatingFuzzTesting,
  FixReverter, TestingNIjfp} fall on a spectrum
from manually sourced to automatically synthesized.
We opt for manual sourcing, allowing us to more readily
maintain \textit{ground truth} and ensure that every mutant
violates some aspect of the property specification.
\name{} supports a terse syntax for incorporating these mutants
into the systems under test.
In \sectionref{sec:populating}, we detail the
systems evaluated in this paper.

\subsubsection{Use minimal, but precise interfaces.}
\label{sec:interfaces}


A key challenge during \name's development is that it needs to gather
data about the testing effectivenes of a wide variety of existing
frameworks, each of which reports such data in an ad-hoc non
standardized manner:
most frameworks only report the number of inputs that were generated
before a counterexample was found; very few offer timing statics; none
offer a detailed breakdown of generation, testing, or minimization
time.
Similarly, most frameworks report the number of inputs that fail to
satisfy a precondition as discards; some (like Rust's QuickCheck and
Racket's RackCheck) do not.
Most frameworks allow for setting a limit on the number of tests; very
few allow for setting similar time limits.
%
%
Moreover, any printing of such data is optimized for human
readability---with little to no consideration to how machine-readable
this output is.
%
%

%
%

To tame this diversity, we settled on a set of metrics for PBT
frameworks that are easily measurable, and on a precise output format
that developers can adhere to: JSON defined using a schema that
frameworks can validate themselves on. We built adaptors to that
schema for frameworks across multiple languages
(\sectionref{sec:cross}).\footnote{\url{https://github.com/alpaylan/etna-cli/blob/main/PROTOCOL.md}}
%


%
%

\subsubsection{Every \name{} capability should be available to use manually.}

\name{} acts as an orchestration mechanism that invokes testing tools,
parses their results, and performs analysis on them. Such
orchestration is, even with the best of intentions, fragile as a
result of loosely coupled independently developed systems working
together. In turn, as we found out from experience, any opaqueness in
this process can result in unrecoverable failures. The final guiding
principle of \name{} is to ensure not just that such opaqueness
doesn't exist, but that users can also reproduce the steps of any
\name{} experiment manually, if they so wish.

\subsection{Terminology}

Our mutation-testing based evaluation is built upon {\bf tasks}: a
mutant-property pair where the mutant causes the property to fail.
As any given program can give rise to multiple tasks~--- it might need
to satisfy multiple properties or be subjected to multiple mutants~--- we
organize tasks into {\bf workloads}. Each workload comes with several
components: data type definitions; variant implementations of
functions using these types; and a property specification of these functions.

We call a \pbt{} paradigm at the level of a library a
  {\bf framework}, which should contain functions for (a) constructing
properties, (b) constructing generators, and (c) running tests. For
instance, QuickCheck, QuickChick, SmallCheck and LeanCheck are all
examples of frameworks. And we call a \pbt{}
paradigm at the level of how to use a framework to write generators a
  {\bf strategy}. Examples of such strategies include type-based random
generation, manually written bespoke generation, or exhaustive
enumeration of the input space.

\subsection{Using \name{}}

\name{} is designed to be an extensible platform that flexibly
accommodates new workloads, strategies, frameworks, and languages,
built with inspiration from modern package managers such as Cargo~[\citeyear{Cargo}].
At its core is an experiment driver that provides three main pieces of
functionality: (a) toggling between variant implementations in a
directory of workloads; (b) compiling and running each strategy on
each task; and (c) analyzing the results.

Users interact with \name{} by creating such {\em experiments}:
projects that can host multiple tests, pull workloads from \name{} and
modify them as the users wish, holding the raw data produced by
experimentation, the analysis results and figures.

%

The users, for instance, can reuse the
\href{http://github.com/alpaylan/etna-exp}{experiment repository}
that holds the tests and experimentation scripts for
this paper and replicate our results. Users can also create a new experiment from
scratch, add existing workload/strategy pairs in the
\href{http://github.com/alpaylan/etna-cli}{\name{} repository} by running the following
sequence of command line interactions:

\begin{bashcode}
  etna experiment new myexp
  etna workload add --experiment myexp --workload rocq bst
  etna experiment run --tests rocq/bst
\end{bashcode}

For each workload, we provide a default test, accessible via
\codeinline{<language>/<workload>}. This test runs all existing
tasks (mutant-property pairs) against all the existing strategies
in the workload. The users can recreate a test file from scratch under
\codeinline{<experiment>/tests} folder that follows the
\href{https://github.com/alpaylan/etna-cli/blob/main/schemas/test.json}{test schema}
for running custom tests.

A common use case for the users is to evaluate new generation strategies for
an existing framework in \name{}. The users can add the new strategy to the workload,
adjust the build instructions to compile this new strategy into a form accessible
from a command line runner, and easily run the default test for obtaining the results
of the evaluation. Adding a new framework requires more work, they need to first
implement an adaptor for the framework that provides the information the CLI requires,
following the existing examples of
\href{https://github.com/alpaylan/etna-cli/blob/docs-bst/workloads/Haskell/config.json}{Haskell},
\href{https://github.com/alpaylan/etna-cli/blob/docs-bst/workloads/Rocq/config.json}{Rocq},
\href{https://github.com/alpaylan/etna-cli/blob/docs-bst/workloads/OCaml/config.json}{OCaml},
\href{https://github.com/alpaylan/etna-cli/blob/docs-bst/workloads/Racket/config.json}{Racket},
or
\href{https://github.com/alpaylan/etna-cli/blob/docs-bst/workloads/Rust/config.json}{Rust}. The
adapters follow a schema available at the \name{}
\href{https://github.com/alpaylan/etna-cli/blob/main/schemas/invoke.json}{repository}.

Finally, to contribute a new workload, users can implement the system
under test just as they would ordinary code in a supported
language. They can then encode mutants via special comment
syntax embedded within the implementation. For example, consider the
following implementation of \verb|insert|, together with a triggerable
bug, in Haskell syntax: 
\begin{hask}
  insert k v Leaf = Node Leaf k v Leaf
  insert k v (Node l k' v' r)
  {-! -}
  | k < k' = Node (insert k v l) k' v' r
  | k > k' = Node l k' v' (insert k v r)
  | otherwise = Node l k' v r
    {-!! insert_duplicate_entries -}
    {-!
      | k < k' = Node (insert k v l) k' v' r
      | otherwise = Node l k' v' (insert k v r)
      -}
    {- !-}
\end{hask}
The correct (i.e. uncommented) implementation of \verb|insert| ensures
that the search tree invariant is maintained: every key in the left
subtree of a node is smaller than its root, and every key in the right
subtree is greater. In specially marked comments, a mutant is
specified, which triggers a bug if enabled by not considering the case
where the key being inserted is already present in the tree.

The mutation syntax is rather straightforward, each variation (a correct \emph{base}
implementation with one or more \emph{mutants} that constitute bugs) starts with
a \codeinline{<comment-begin><marker> <comment-end>}, and ends with \codeinline{<comment-begin> <marker><comment-end>},
parametrized based on the language. The beginning marker is followed by some piece of plain
code that is the base implementation, which is followed by a sequence of mutations,
a header in the form of \codeinline{<comment-begin><marker>\{2\} <name> <comment-end>},
and a body enclosed in \codeinline{<comment-begin><marker> <body> <comment-end>}. The EBNF
grammar for the syntax can be found in the repository for the mutation injection tool~\cite{marauders}
we implemented as a complement to \name.

\pagebreak
\subsection{Analysis and presentation}

\begin{wrapfigure}{r}{.5\textwidth}
  \centering
  \includegraphics[width=0.5\textwidth]{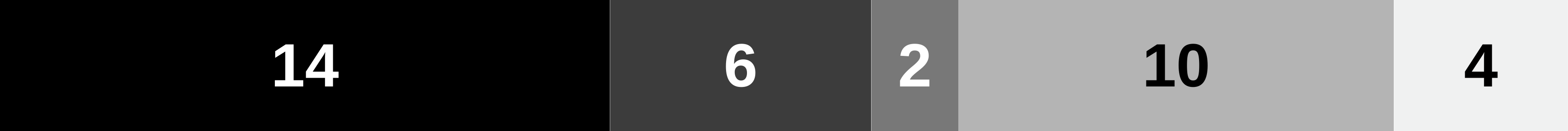}
  \\[2ex]
  \includegraphics[width=0.5\textwidth]{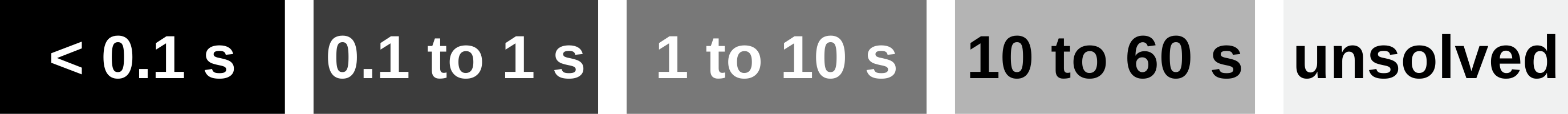}
  \vspace*{-2em}
  \label{fig:visual-example}
\end{wrapfigure}
Though \name{} supports customizable experiments, we choose a standard
set of defaults for the experiments in this paper.  We run each strategy on
each task for a set amount of trials (10 unless otherwise specified) and with a
set timeout (60 seconds). We then measure if the
strategy was able to {\em solve the task}, i.e. find the injected bug in all
trials within the given time frame. Multiple trials account for the non-determinism
of random generation strategies, and results are simple averages unless indicated otherwise.

Our first attempts at presenting this data were hard to interpret: what does it
mean, for example, if one strategy takes an average of two seconds and the other
an average of three? Rather than present a slew of raw numbers, we wanted a data
representation that captures a user's experience of interacting with \pbt{}
tools, so that visual differences in the representation correspond to tangible
differences in performance.  The figure above demonstrates our
solution: a {\em task bucket} chart.
For every strategy we classify tasks ranging from ``solved instantly'' to
``unsolved'', depicted with progressively lighter shades. For example,
for the strategy/workload combination in the figure,
14 tasks are solved very quickly (the darkest shade) while four are
not solved at all (the lightest).

In case a task bucket chart does not show enough detail, especially in
head-to-head comparisons, we
also support statistical analyses like Mann--Whitney U tests\footnote{The
  Mann--Whitney U test is a nonparametric test that compares data samples from two
  different distributions. We use it here because it makes no assumptions about
  the distributions being compared.}
(see \sectionref{sec:comparing-frameworks}).
%
%
%
%

\section{Populating the platform}\label{sec:populating}

We have integrated a number of PBT frameworks and workloads into
\name, both for our own use in \sectionref{sec:haskell} -
\sectionref{sec:cross} and for potential users to use and compare
against.

\subsection{Languages and frameworks}\label{sec:languages-and-frameworks}

Haskell is an obvious starting point: as the language that hosts
QuickCheck, it is the lingua franca of PBT research.  We focus on
three Haskell frameworks: QuickCheck, of course;
SmallCheck~\citep{RuncimanNL08}, a competitor to QuickCheck that does
enumerative testing; and LeanCheck~\citep{RudyMatelaThesis}, a more
modern enumerative framework.

Our second language of choice is Rocq.  While \has{} is blessed with
many \pbt{} frameworks, PBT in \rocq{} is built on a single framework:
QuickChick~\citep{Pierce:SF4}. However, QuickChick is a rich ecosystem
that supports a variety of different strategies for input
generation~\citep{LeoThesis2018,GeneratingGoodGenerators,FuzzChick},
so there is plenty to study and compare.

The third language we focus on is OCaml. Similar to the Haskell ecosystem,
OCaml users can reach for a variety of random testing frameworks in the OCaml ecosystem,
from QuickCheck variants such as QCheck~\citep{QCheckOCaml} or
base\_quickcheck~\citep{base_quickcheck}, to AFL~\citep{zalewski2015afl-github} powered fuzzers like
Crowbar~\citep{Crowbar}.

For these three languages, we perform intra-language experiments comparing
different generation strategies (\sectionref{sec:haskell}, \sectionref{sec:rocq},
\sectionref{sec:ocaml}).
We will also show how to perform cross-language experiments with \name{}
(\sectionref{sec:cross}), using strategies both from these languages,
as well as Racket (using its RackCheck framework~\citep{rackcheck})
and Rust (using~\citep{quickcheck-rs}).
%
\name's extensible design means that adding new languages is straightforward; we
discuss languages that we plan to add to the platform in
\sectionref{sec:relwork}.

\subsection{Workloads}\label{sec:workloads}

Our initial set of workloads is drawn from three application domains
that are of practical interest to the functional programming community
and that have featured prominently in the \pbt{} literature. These
workloads feature in the following sections' experiments, although not
every workload is used for every experiment. A detailed description of
each workload, together with a list of properties and associated
mutants, can be found in the
\href{https://github.com/alpaylan/etna-cli/tree/docs-bst/docs/workloads}{repository}.

\newcommand{\workloadarea}{\ourparagraph}
\newcommand{\workload}[1]{}

\workloadarea{Data structures}
\workload{Binary Search Trees}
The first workload focuses on a functional data structure that is
ubiquitous in the literature: binary search trees. Multiple
PBT papers have focused on BST
generation, including John Hughes's {\em How to Specify
It!} (\citeyear{Hughes2019HowTS}), an extended introduction to specifying
properties using QuickCheck. Our \namedsystem{BST} workload ports the
mutations and properties from that paper.
\workload{Red-Black Trees}
The second workload focuses on another popular functional data
structure, red-black trees, including self-balancing insertion and
deletion operations that are notoriously easy to get wrong.
RBTs have also been studied in the PBT
literature~\citep{Luck,RuncimanNL08,MistaR19B,Klein09}. Our \namedsystem{RBT}
workload combines the \namedsystem{BST} mutants with additional mutants that
focus on potential mistakes when balancing or coloring the tree.

\workloadarea{Lambda calculi and type systems}
\workload{Simply Typed Lambda Calculus}
The third workload centers around a DeBruijn index based
implementation of the simply typed lambda calculus with
booleans. Bespoke generators for producing well-typed
lambda terms is a well studied problem in the
literature~\citep{PalkaAST11, MidtgaardNPFH17}, while the mutations for
STLC included in our case study are drawn from the appropriate
fragment of a System F case study~\citep{CombinatorialPBT},
dealing mostly with mistakes in substitution, shifting, and lifting.
\workload{System \FSUB}
For a more complicated fourth workload \FSUB{} revolving around calculi
and type systems, we turn to the full case study
of~\citet{CombinatorialPBT} and extend it with subtyping. This
allows for significantly more complex errors to be injected (such as
those dealing with type substitution, shifting, or lifting).
Bespoke generators for System F have been the subject of recent
work~\citep{CombinatorialPBT, BlockchainTesting} and can be
straightforwardly extended to handle subtyping.
The fifth workload involves a parser and pretty-printer for Lu, a
language based on Lua; the implementation of Lu was drawn from a Haskell course
at the University of Pennsylvania. We specify correctness through a round-trip
property: printing a valid Lu expression and then parsing it should result in
the original expression. We release the Lu parser workload with an accompanying
bespoke generator.

\workloadarea{Security}
\workload{Information Flow Control}
The sixth and final workload focuses on a security domain: information
flow control. The IFC case study, introduced
by~\citet{TestingNoninterferenceQuickly, TestingNIjfp}, explores the
effectiveness of various bespoke generators for testing whether
low-level monitors for abstract machines enforce noninterference:
differences in secret data should not become publicly visible through
execution. Violations in the enforcement policies are introduced by
systematically weakening security checks or taint propagation rules,
exploring all possible ways of introducing such violations.

\section{Experiments: Haskell}\label{sec:haskell}

We next report on our experience using
\name{} to probe different aspects of testing effectiveness. Our
first set of observations are on the PBT frameworks and strategies
available in Haskell.

\subsection{Comparing frameworks}\label{sec:comparing-frameworks}

In the first experiment, we assess the ``out of the box'' bug-finding
abilities of three Haskell frameworks~--- \qc{},
\namedsystem{SmallCheck}, and
\namedsystem{LeanCheck}.
We examine four strategies. For the \textit{bespoke} strategy, we
manually write a \qc{} generator that always produces test inputs that
satisfy the property's precondition. This serves as a ``topline'' for
the other strategies: a high-effort generator that solves all of the
tasks easily.
The other three strategies~--- one per framework~--- are all
\textit{naive}.
The \qc{} strategy uses the
\namedsystem{generic-random} library to derive its generator
automatically, with constructors chosen at each step with
uniform probability and a size parameter that decreases on
recursive calls to ensure termination.
For the enumerative frameworks, \namedsystem{SmallCheck} and
\namedsystem{LeanCheck}, we use combinators that follow the type structure.

\begin{figure}[tb]
  \centering
  \begin{subfigure}{\bargraphwidth}
    \includegraphics[width=\textwidth]{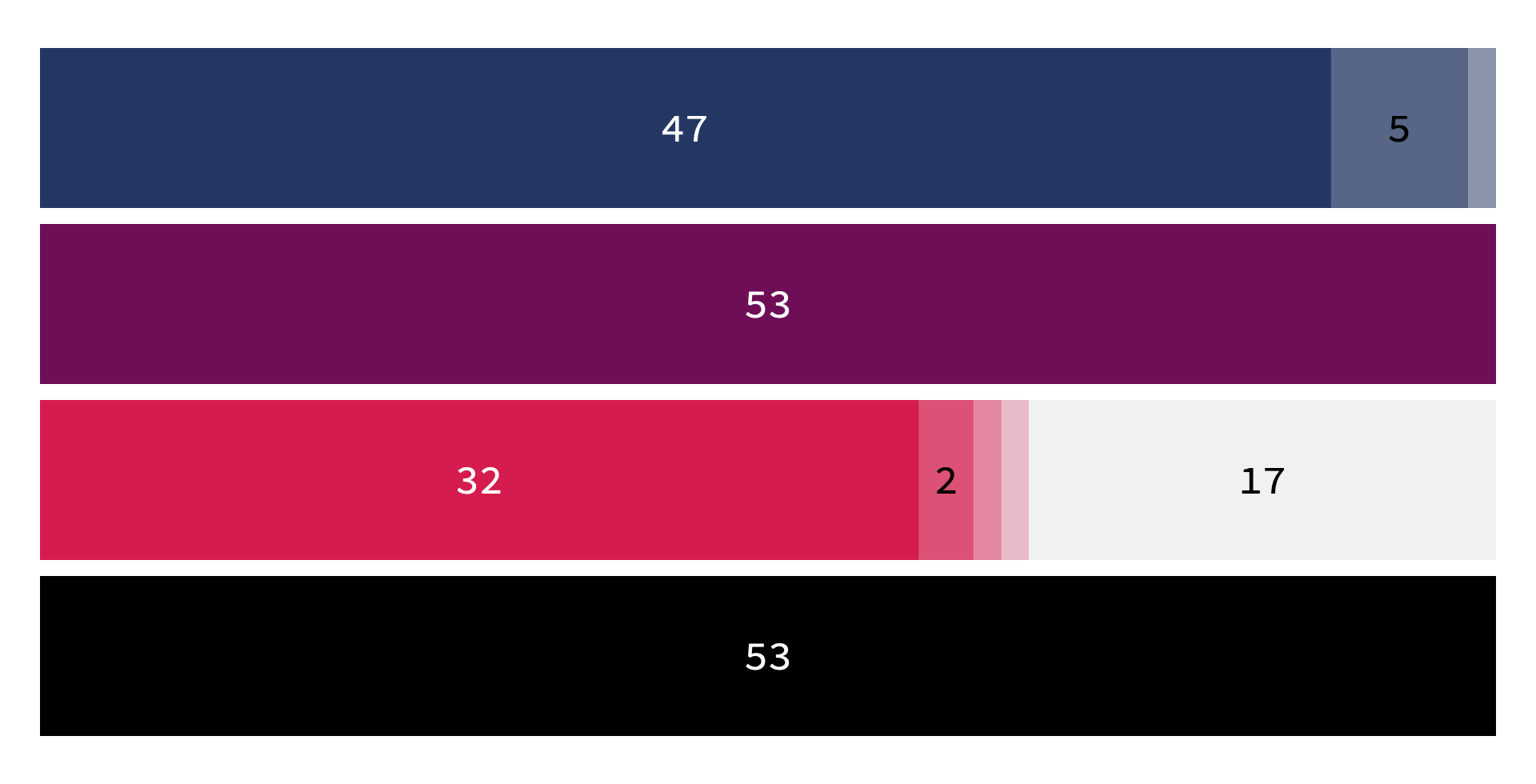}
    \vspace*{-5ex}
    \caption{BST}
  \end{subfigure}
  \begin{subfigure}{\bargraphwidth}
    \includegraphics[width=\textwidth]{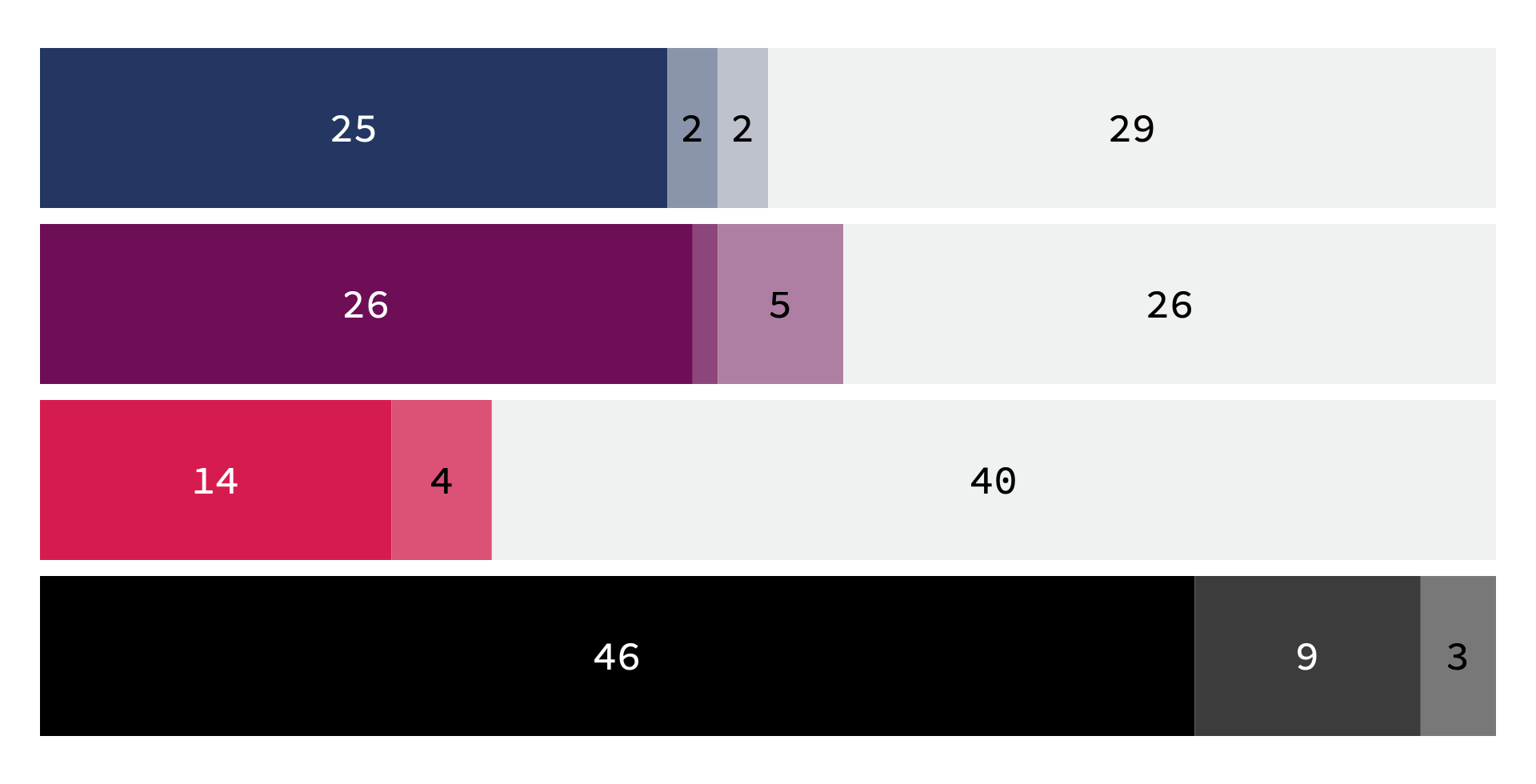}
    \vspace*{-5ex}
    \caption{RBT}
  \end{subfigure}

  \begin{subfigure}{\bargraphwidth}
    \includegraphics[width=\textwidth]{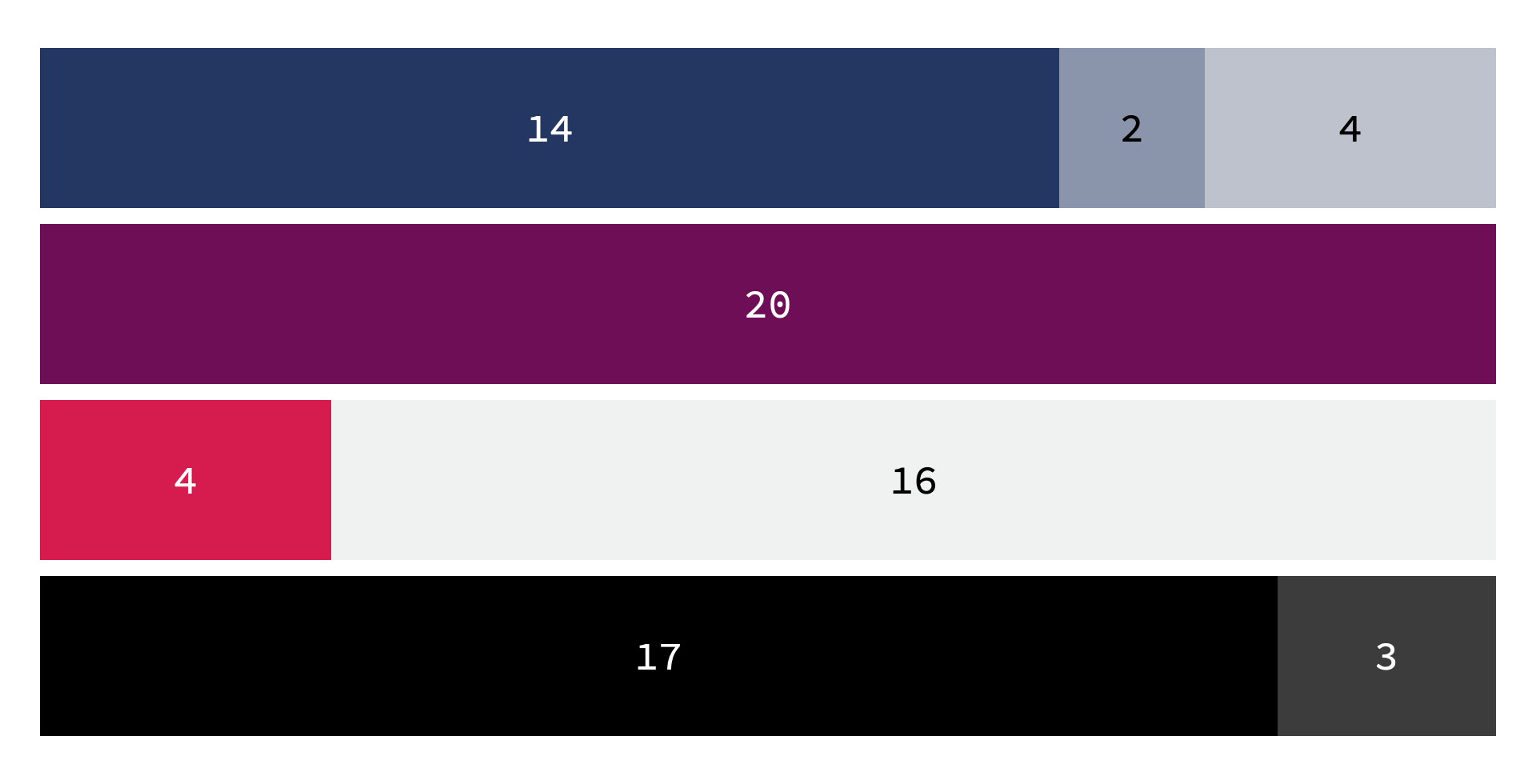}
    \vspace*{-5ex}
    \caption{STLC}
  \end{subfigure}
  \begin{subfigure}{\bargraphwidth}
    \includegraphics[width=\textwidth]{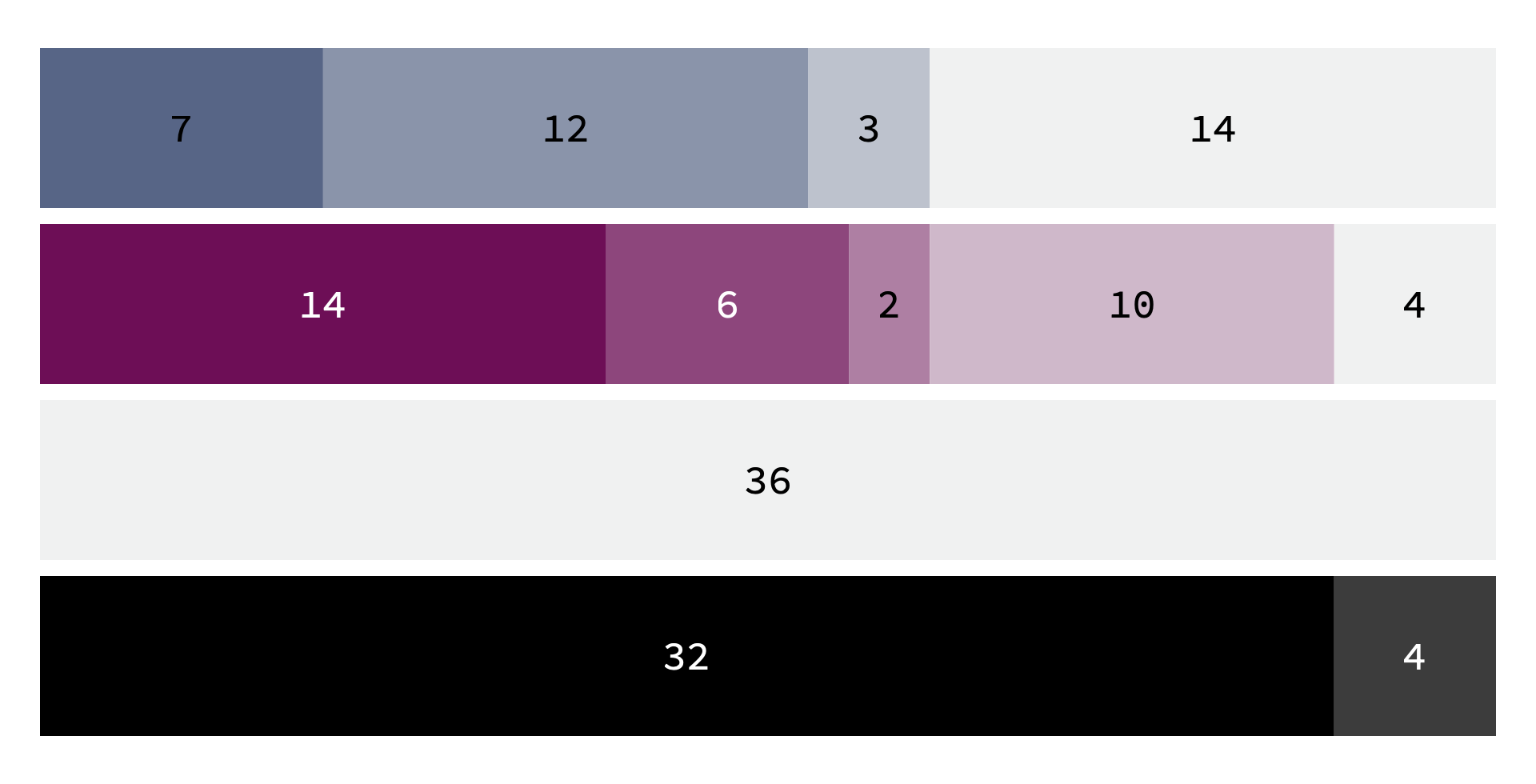}
    \vspace*{-5ex}
    \caption{\FSUB}
  \end{subfigure}

  \captionsetup{justification=centering}
  \caption{
    Effectiveness of Haskell generation strategies on four
    workloads.\\
    \explaincolor{chart_blue}{naive QuickCheck},
    \explaincolor{chart_purple}{naive LeanCheck},\\
    \explaincolor{chart_pink}{naive SmallCheck},
    \explaincolor{chart_black}{bespoke QuickCheck}.
  }
  \label{fig:haskell1}
\end{figure}

We evaluate these strategies against four workloads: Binary Search
Tree (BST), Red-Black Tree (RBT), Simply-Typed Lambda Calculus (STLC),
and System-F with Subtyping (\FSUB). The results of this experiment
are visualized in Figure~\ref{fig:haskell1}. Our key take-aways are
discussed below.

\textit{The bespoke strategy outperforms the naive strategies along
  multiple
  axes.}  For example, looking at the naive \qc{} strategy (the
others are similar), the bespoke strategy solved all tasks, while the
naive strategy failed to solve 43 tasks. Among tasks that both strategies
solved, using a Mann--Whitney U test with $\alpha = 0.05$, we find that the
bespoke strategy's average time to solve a task was (statistically)
significantly lower
in 83 out of 124 tasks and the average valid inputs to solve a task were lower
for 89 out of 124
tasks.
That is, the bespoke strategy
found more bugs, more quickly, and with better quality
tests.

\textit{Between the two enumeration frameworks, \namedsystem{LeanCheck}
  substantially
  outperforms \namedsystem{SmallCheck}} on these workloads.
\namedsystem{LeanCheck} had an 82\% solve rate, while
\namedsystem{SmallCheck}'s was only 35\%.
On one \namedsystem{BST} task, \namedsystem{LeanCheck} found the bug in about
a hundredth of a second on average, while \namedsystem{SmallCheck} required 26
seconds.
One reason for these differences may be that \namedsystem{SmallCheck}
attempts to enumerate larger inputs much earlier. In the first thousand binary trees,
\namedsystem{SmallCheck} produces trees with up to ten nodes, while
\namedsystem{LeanCheck} only reaches four nodes. Unsurprisingly, it is
harder for larger trees to satisfy the BST invariant~--- only 1\%
of these thousand \namedsystem{SmallCheck} trees are valid, compared to 13\% of the \namedsystem{LeanCheck} trees.
And across all workloads, we can calculate the rate at which they enumerate test inputs, by
aggregating over the tasks that they both solved and dividing by the total number of tests by the
total time spent. We find that \namedsystem{LeanCheck} produces over a hundred
times more tests per second than \namedsystem{SmallCheck}.

\textit{\namedsystem{LeanCheck} also outperforms naive \qc{}.}
It is illuminating to consider failed tasks that were \textit{partially solved}:
the bug was found in at least one trial and not found in at least one trial. There is one such task for \namedsystem{LeanCheck} and 14 for \namedsystem{QuickCheck}. For \namedsystem{LeanCheck}'s partially solved task, the inputs required are the same for each trial, but the time fluctuates between 55 and 65 seconds. That is, this is a situation where a task nears~--- and sometimes exceeds~--- what \namedsystem{LeanCheck} can reach with the one minute time limit.
\namedsystem{QuickCheck}'s partially solved tasks are also interesting. Of the 13  that \namedsystem{LeanCheck} solves but \namedsystem{QuickCheck} does not, 10 are partially solved by \namedsystem{QuickCheck}. This suggests that there are situations where a deterministic approach may be more reliable than a random alternative: \namedsystem{LeanCheck} solves these tasks consistently and relatively quickly, while \namedsystem{QuickCheck} sometimes takes less than a second, sometimes nearly a minute, and sometimes times out.

Similarly, in \namedsystem{STLC}, naive \namedsystem{LeanCheck} solves three more tasks within the first bucket than the bespoke strategy. Upon closer inspection, these are tasks that the bespoke strategy sometimes solves in under 100 inputs but sometimes requires over 10,000 inputs, leading to an average slightly above the 0.1 second threshold; as before, \namedsystem{LeanCheck} does not experience this variability.

\textit{A note about memory usage.} \namedsystem{LeanCheck} is documented\footnote{\url{https://github.com/rudymatela/leancheck/blob/master/doc/memory-usage.md}} to be memory intensive, especially when run for prolonged periods of time, as we do here. Our experiments using \namedsystem{LeanCheck} were conducted on a server with plenty of memory, allowing us to complete trials without issues. Future work might consider the relative space complexities of different frameworks.

\subsection{Exploring sized generation}\label{sec:haskell:size}

We next explore the sensitivity of bug-finding to
various parameters, starting with input size.

A significant part of generator tuning is ensuring that the
generated inputs are well sized. Conventional wisdom in random
testing posits that there is a ``combinatorial advantage'' to testing with large
inputs, since they can exercise many program
behaviors at once; tools like {\em
    QuickCover}~\citep{CombinatorialPBT} capitalize on this
notion to make testing more efficient. But are large inputs {\em
    always} better? We used our BST workload to investigate.

We conducted this experiment on the \qc{} framework, using a bespoke
strategy to focus attention on the quality of
the distribution of \textit{valid} inputs.
We used a generator from~\citet{Hughes2019HowTS}, which
generates a list of keys and then inserts each key into the tree, because it
gives precise control over final tree sizes. We choose the keys for a $n$-node
tree from a range of integers $1$ to $2n$. This range is large enough to allow for
sufficient variety in shape and content but not so large that a randomly generated key
in this range is unlikely to be in the tree.

We then measured the bug-finding effectiveness of the generator at different sizes $n$. Thanks to \name's flexibility, we could vary the size in the script and otherwise treat this experiment
as we would any other where we wanted to compare several strategies.

\begin{wrapfigure}{r}{.5\textwidth}
  \includegraphics[width=0.5\textwidth]{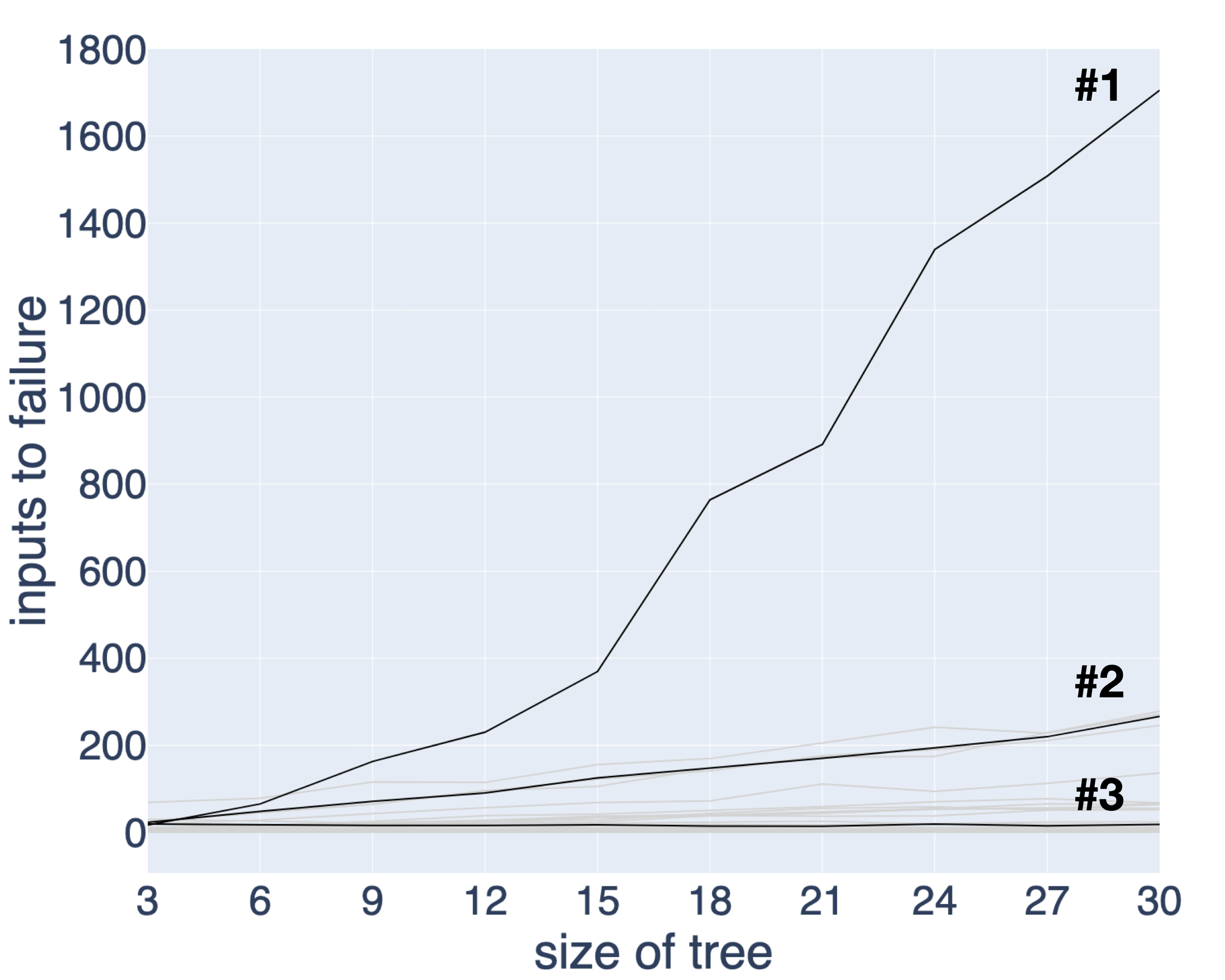}
  \caption{Number of generated inputs
    (averaged over 100 trials) to solve each \namedsystem{BST}  task,
    as input size increases from three to
    30 nodes.}\label{fig:haskell2}
  \vspace*{-2ex}
\end{wrapfigure}

\ourparagraph{Results} Figure~\ref{fig:haskell2} plots the size of the tree versus the average number of inputs to solve a
task; each trace represents one task.  Some noteworthy
traces, highlighted in black, are discussed below.

\textit{Larger trees can be {\em worse} for bug-finding, for properties
  that rely on dependencies between their inputs.} We
found that, for BST, small trees were generally sufficient to
find bugs, and performance got significantly worse for
some tasks as trees got larger.

For example, task \#1, which has the steepest upward curve,
involves a mutant where the \codefont{delete} function fails to remove a key
unless that key happens to be the root.
The property takes one tree and two keys as
inputs and checks that removing the keys in either order leads to the same result.
Together, these mean that the task is only solvable when one key \codeinline{k} is the root of the tree and the other key \codeinline{k'} becomes the root after
deleting \codeinline{k}.
The probability of satisfying this condition decreases as the size of the tree
increases, so larger trees take more inputs to solve this task.

Task \#2 is a similar story.
It takes a tree and two key-value pairs; this time, the task is only
solvable when the two keys are the same (and the two values are different),
a probability that is inversely
proportional to the size of the tree. These two tasks demonstrate situations
where the inputs to a property need to be related in a mutant-specific way,
and large trees are less likely to satisfy this dependency relationship.

\textit{Not all tasks with dependencies between their inputs are harder
  to solve with larger trees.}
Unlike \#1 and \#2, the curve for task
\#3 is mostly flat, even though it has a similar dependency.
The mutant here causes the \codeinline{union}
operation to fail by
occasionally preferring the wrong value if both trees
contain the same key; the property takes a key \codeinline{k} and two trees and
checks that \codeinline{k} exists in the union of the trees when it
exists in either tree.
Since this mutant causes problems with keys that appear in both trees, the
property only fails when \codeinline{k} is in the input trees. That is,
there {\em is} a dependency between the inputs, but this dependency
does {\em not}
scale with the size of the tree.

\ourparagraph{Discussion}
We have seen that larger inputs sometimes not only fail to provide a
combinatorial advantage but in fact can provide a dependency disadvantage.
The size of the main input~--- in this case, the tree~---
cannot be evaluated in a vacuum. Instead, the particulars of the mutant and
property can lead to dependencies between the property inputs that must be
satisfied in order to detect the mutant.
Our size exploration is thus a cautionary tale: PBT users should not
naively expect that larger inputs are better, especially for properties with multiple inputs.

This exploration suggests a few recommendations for improving both testing
frameworks and individual users' choices of properties.
  {\em (1) Do not treat property inputs as independent.}  The
difficulties with the above properties arise, in part, because
\namedsystem{QuickCheck} automates generation of multiple
inputs by assuming that each input can be generated independently~--- but
treating inputs independently can lead to unintuitive testing
performance.  Frameworks like Hedgehog explicitly avoid introducing a
generator typeclass so as to force users to build generators by hand;
our results lend credence to that design choice.
  {\em (2) Think carefully about properties with multiple inputs.}
Testers should prefer
properties with fewer inputs where possible.
When this is infeasible,
testers should think carefully about potential interactions between
their property's inputs and write generators that take those
interactions into account.

\subsection{Enumerator sensitivity}\label{enums}
Papers about enumeration frameworks sometimes speak of enumeration as a kind
of exhaustive
testing~--- validating the program's behavior within a ``small
scope''~\citep{SmallScope}. But realistic testing
budgets often mean that exhausting all inputs up to some interesting size or
depth is not possible: enumeration is {\em expensive}. Thus, the actual
performance of
enumeration frameworks like \namedsystem{SmallCheck} and \namedsystem{LeanCheck} is impacted by the specific order in
which values are enumerated. In this section we examine
some factors that,
perhaps unexpectedly, impact bug-finding performance.

There are many axes along which order could vary. We have explored
two: the order of the inputs to each property and the order of
constructors in an algebraic data type.
%
%
We conduct this experiment on \namedsystem{SmallCheck} and
\namedsystem{LeanCheck}, using the \namedsystem{BST} and
\namedsystem{RBT} workloads where many of their properties have
multiple inputs, including a combination of \codeinline{Tree}s and
\codeinline{Int} keys.  One enumeration strategy uses the default
properties, with the trees passed in first, and one uses properties
with the trees last~--- for example, \codeinline{(Tree, Tree, Int)} vs.
\codeinline{(Int, Tree, Tree)}.

\begin{figure}
  \centering
  \begin{subfigure}[t]{\bargraphwidth}
    \includegraphics[width=\textwidth]{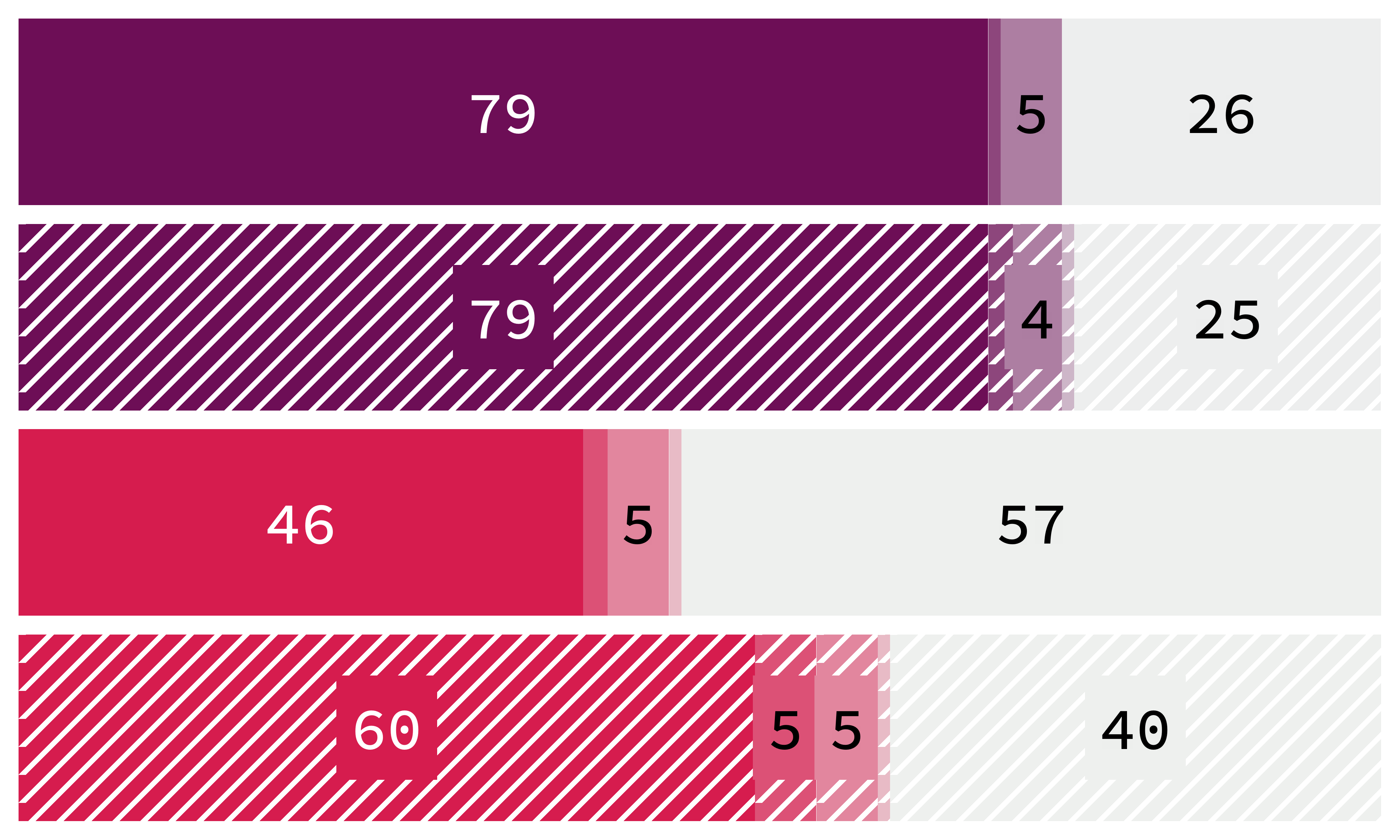}
    \vspace*{-4ex}
    \caption{Enumerator performance on the BST and RBT workloads when
      the trees are at the start of the properties (top rows) versus
      when they are at the end (bottom rows).}
  \end{subfigure}
  \quad
  \begin{subfigure}[t]{\bargraphwidth}
    \includegraphics[width=\textwidth]{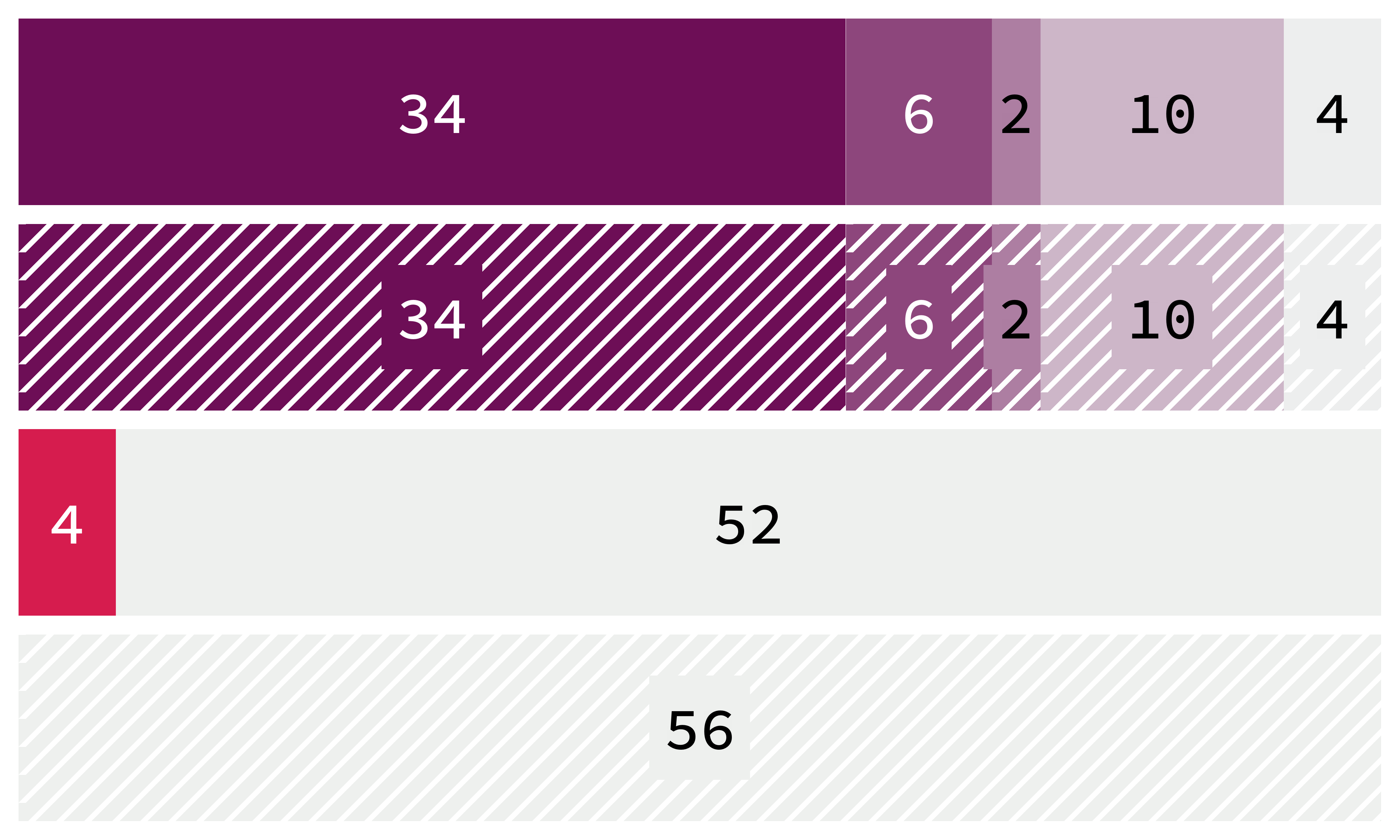}
    \vspace*{-4ex}
    \caption{ Enumerator performance on the STLC and FSUB
      workloads when the constructor enumeration order aligns
      with the definition of the data type (top rows) versus
      when the orders are reversed (bottom rows).}
  \end{subfigure}
  \caption{
    \explaincolor{chart_purple}{naive LeanCheck (default order)},
    \explaincolorhatched{chart_purple}{naive LeanCheck (reverse order)},\\
    \explaincolor{chart_pink}{naive SmallCheck (default order)},
    \explaincolorhatched{chart_pink}{naive SmallCheck (reverse order)}.
  }
  \label{fig:enum-sense}
\end{figure}

\ourparagraph{Results} We count the number of tasks that are solved by
the same framework under each of the two orderings. The results
are shown in Figure~\ref{fig:enum-sense}.

For \namedsystem{LeanCheck}, the tree-last strategy solved one
additional task that the tree-first strategy did not (completing in
about 38 seconds instead of timing out at 60).
For \namedsystem{SmallCheck}, the tree-last strategy solved 17 more
tasks than tree-first, taking between 0.002 and 7 seconds. The low end is
especially remarkable: simply by enumerating \codeinline{(Int, Tree, Tree)}s rather than
\codeinline{(Tree, Tree, Int)}s, \namedsystem{SmallCheck} finds a counterexample almost
immediately instead of timing out.

\ourparagraph{Discussion} A deeper dive into the enumeration
frameworks to explore
these differences fully would be worthwhile, but what jumps out
even from these simple experiments is the question of
\textit{sensitivity}. We see that the trees (BST/RBT) are much more sensitive
to enumeration order than the languages (STLC/FSUB) in our experiments.
The potentially pivotal role
of enumeration order in the success or failure of these strategies means
that users of these enumerative frameworks need to be careful of configuration
settings that would be immaterial in their random counterparts.
As a meta point, we put the tree data types at the front of each
property as a matter of convention; it was not until much later that
we realized the inadvertent effect on the performance of the
enumerators!

\section{Experiments: Rocq}\label{sec:rocq}

After focusing on the multi-framework landscape of Haskell in the
previous section, we now turn our attention to the single-framework
but multi-strategy landscape in Rocq.
As discussed in
\sectionref{sec:languages-and-frameworks}, PBT in Rocq revolves
around QuickChick~\citep{LeoThesis2018}, which, in addition to the
type-based and bespoke strategies that we explored in Haskell,
provides two additional options: a {\em specification-driven} strategy
that derives correct-by-construction generators from preconditions in
the form of inductive relations~\citep{GeneratingGoodGenerators} and a
  {\em type-driven fuzzer} strategy that combines type-based generation with mutation
informed by AFL-style branch coverage to guide the search
toward interesting parts of the input space~\citep{FuzzChick}.

Both papers exemplify the lack of performance
comparisons across approaches discussed in the
introduction. First, \citet{GeneratingGoodGenerators} is evaluated in
a toy IFC example, where only the throughput of
generators is measured against that of a bespoke generator; there is
no measurement of the effectiveness of the strategy in finding bugs.
On the other
hand, FuzzChick~\citep{FuzzChick} is evaluated in the more realistic IFC
workload of \citet{TestingNIjfp} that we will reuse later in this
section, with systematically injected mutations that break the
enforcement mechanism of a dynamic monitor. Still, multiple aspects
of their strategies were left unevaluated, including their performance
on any other workload.

\subsection{Comparison of fuzzers, derived generators, and handwritten generators}
\label{sec:rocq:fuzz}

We aim to fill the evaluation gaps described above. How do
QuickChick's newer strategies compare with the more established
bespoke and type-based ones?  In particular, are they effective at
uncovering bugs across disparate workloads?

We again use the BST, RBT, and STLC workloads, along with a more
complex case study, IFC, pulled from the FuzzChick paper. For the first three case
studies, inductively defined specifications are widely available
(e.g. in Software Foundations~\citep{Pierce:SF}); for IFC, such
specifications do not exist, so the specification-driven generators
of~\citeauthor{GeneratingGoodGenerators} do not apply.

\ourparagraph{Results}
\begin{figure}[tb]
  \centering
  \begin{subfigure}{\bargraphwidth}
    \includegraphics[width=\textwidth]{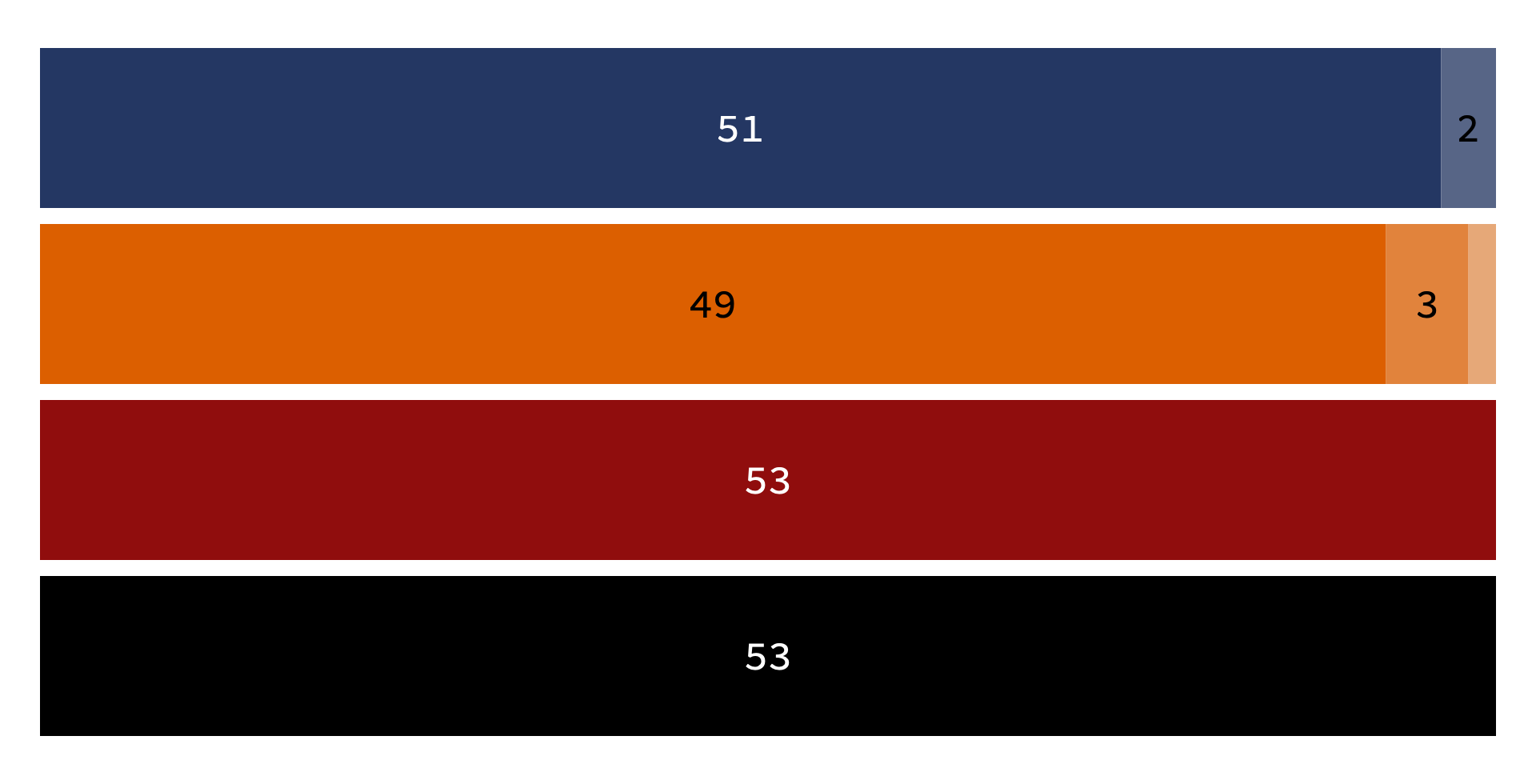}
    \vspace*{-5ex}
    \caption{BST}
    \label{subfig:coq1:bst}
  \end{subfigure}
  \begin{subfigure}{\bargraphwidth}
    \includegraphics[width=\textwidth]{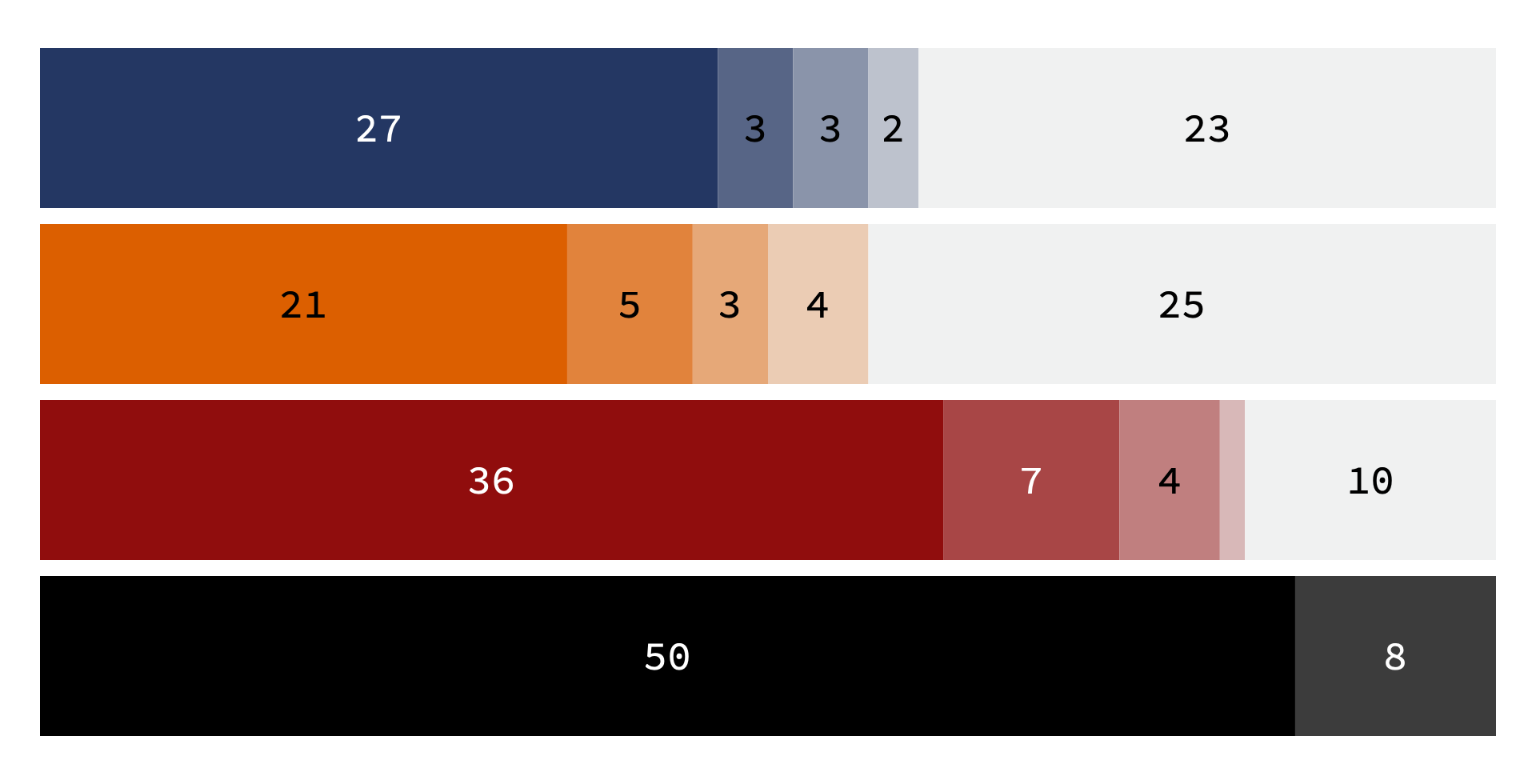}
    \vspace*{-5ex}
    \caption{RBT}
    \label{subfig:coq1:rbt}
  \end{subfigure}

  \begin{subfigure}{\bargraphwidth}
    \includegraphics[width=\textwidth]{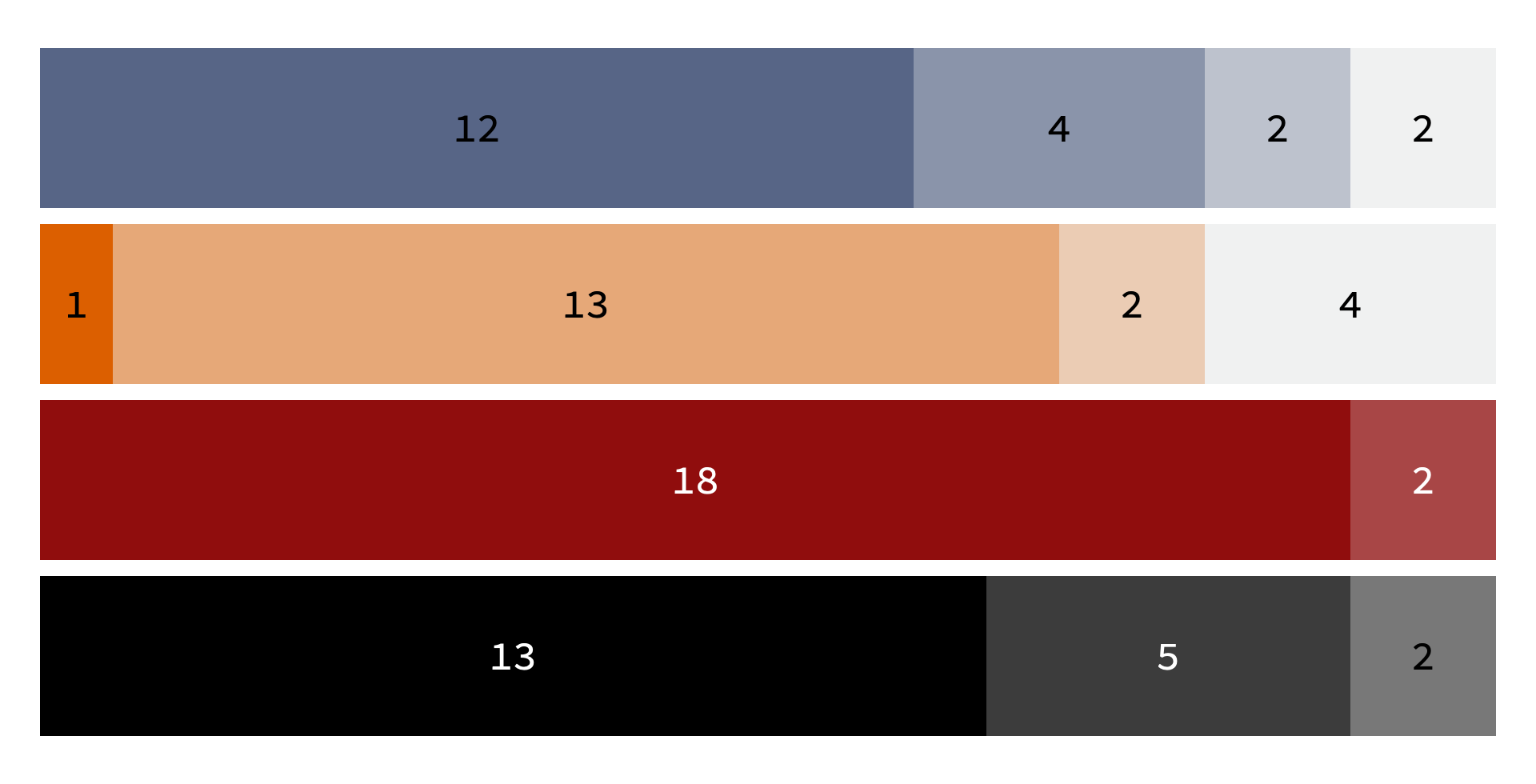}
    \vspace*{-5ex}
    \caption{STLC}
    \label{subfig:coq1:stlc}
  \end{subfigure}
  \begin{subfigure}{\bargraphwidth}
    \includegraphics[width=\textwidth]{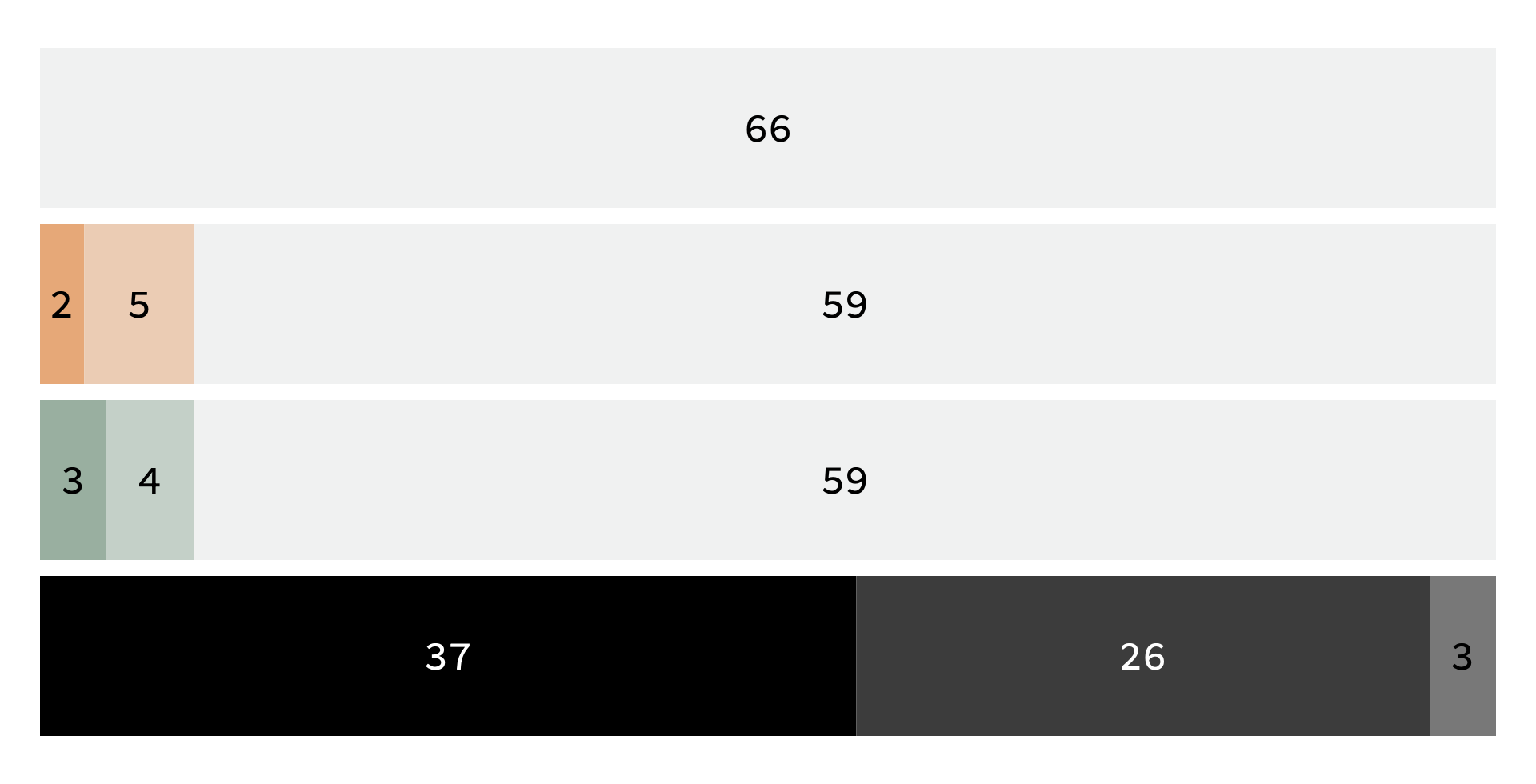}
    \vspace*{-5ex}
    \caption{IFC}
    \label{subfig:coq1:ifc}
  \end{subfigure}

  \captionsetup{justification=centering}
  \caption{Effectiveness of Rocq generation strategies on four workloads.\\
  \explaincolor{chart_blue}{type-based generator},
  \explaincolor{chart_orange}{type-based fuzzer},\\
  \explaincolor{chart_red}{specification-based generator ((a) - (c) only)},
  \explaincolor{chart_green}{variational fuzzer} ((d) only),\\
  \explaincolor{chart_black}{bespoke generator}.
  }
  \label{fig:coq1}
\end{figure}
In Figure~\ref{fig:coq1}, we visualize the results of the experiments
with a task bucket chart.
Results for the
simple BST workload (Figure~\ref{subfig:coq1:bst}) establish a
baseline level of confidence for all four methods, as they are all
able to solve most tasks quickly. Indeed, most of the tasks are solved
by all methods within 0.1 seconds (the darkest color), with the
exception of the \emph{type-based fuzzer}, which falls short on a
few tasks.

\result{1}{Specification-derived strategies are on par with bespoke ones.}
In the harder RBT workload, with its much more complex invariant,
there is a clear performance gap between type-driven strategies
(type-based generator and
type-based fuzzer) and precondition-driven methods
(specification-based generator and bespoke generator).
Precondition-driven methods are able to solve more tasks under 0.1 seconds than
type-driven methods are able to solve within a 60 second timeout. The type-based
generator fails to solve 23 tasks, and the type-based fuzzer fails to solve 25.
The bespoke generator solves all tasks in under ten seconds, and the
specification-based generator solves all but 10 tasks.
We see a similar pattern in the STLC workload, with the precondition-driven
methods outperforming the type-driven ones.

\result{2}{Fuzzers exhibit more variance but outperform type-driven methods for sparse preconditions.}
For the IFC workload, the only precondition-driven strategy is the
bespoke generator, which emerges as a clear winner: noninterference is
a property with a {\em very} sparse precondition, and type-based
methods are basically unable to generate valid inputs. For this
particular workload, we included another fuzzing variant, \emph{variational fuzzer}, borrowed from
the original paper that introduced FuzzChick~\citep{FuzzChick} to
strengthen the connection to the existing literature: rather than
generating a pair of input machines completely at random and then
fuzzing the pair (as in the type-based fuzzer approach), we generate
one input machine and copy it to create a pair that is
indistinguishable by default. The two fuzzers, \emph{type-based
  fuzzer} and \emph{variational fuzzer}, have a clear advantage over the
pure type-based generation approach: the ability to guide generation
allows fuzzers to discover parts of the input space that naive
type-based generation are simply unable to reach.

\begin{wrapfigure}{r}{.5\textwidth}
  \includegraphics[width=0.5\textwidth]{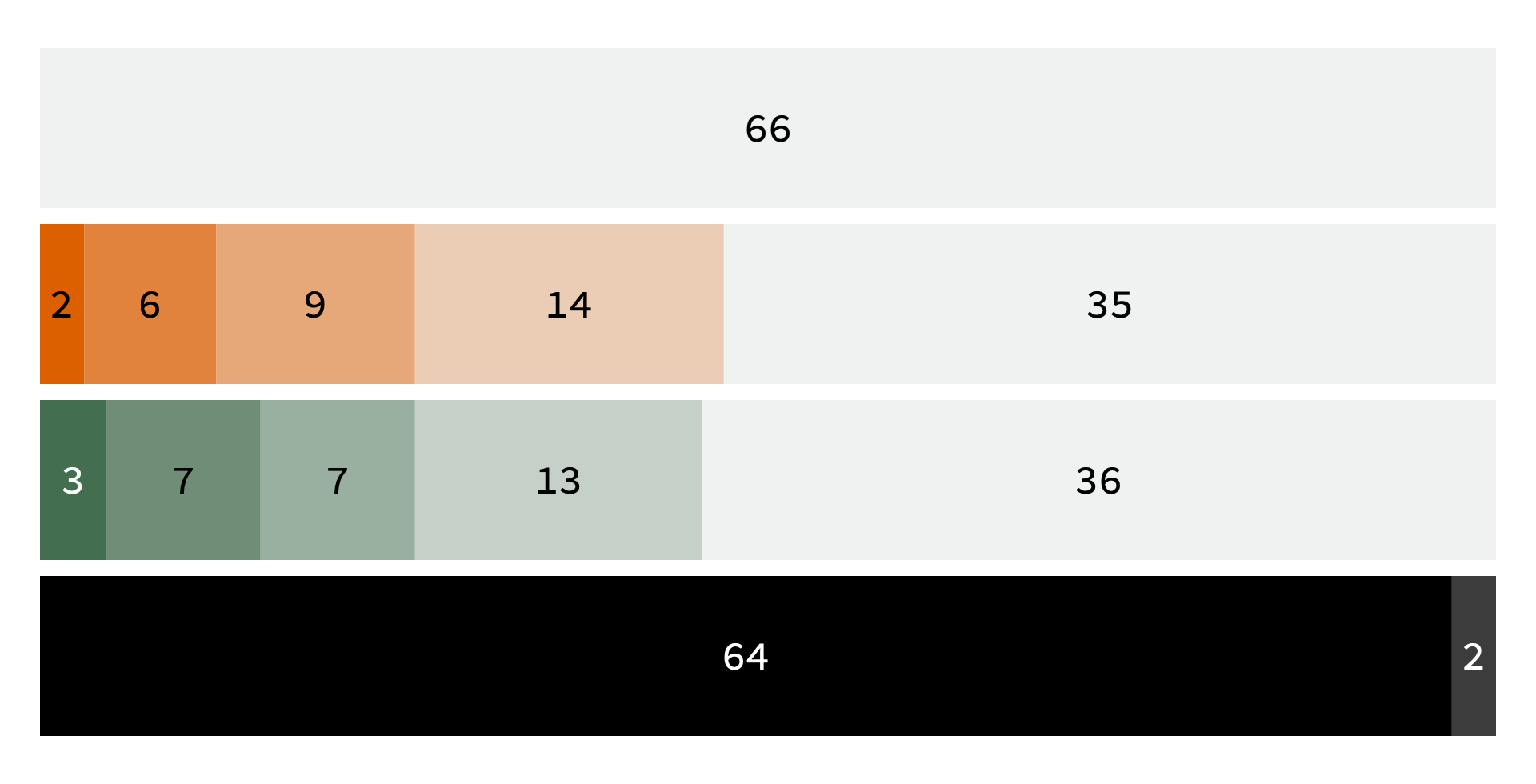}
  \vspace*{-3ex}
  \captionsetup{justification=centering}
  \caption{IFC tasks solved within the timeout in one or more trials.\\
    Empty = type-based generator.\\
    \explaincolor{chart_orange}{type-based fuzzer},\\
    \explaincolor{chart_green}{variational fuzzer},\\
    \explaincolor{chart_black}{bespoke generator}.
  }
  \label{fig:IFC-fine}
\end{wrapfigure}

Yet fuzzers are not reliable in this sense, as
Figure~\ref{fig:IFC-fine} shows: if we include \textit{partially solved}
tasks, fuzzers outperform their generator counterparts.
This further clarifies the picture painted by the first set of
comparisons. Fuzzers may get stuck following program paths that will
not lead to interesting revelations, but sometimes discover paths that
a traditional type-based generator could never hope to reach. In
particular, roughly 30 tasks are solved {\em at least once} through 10
runs (Figure~\ref{fig:IFC-fine}), but less than 10 tasks are
fully solved (Figure~\ref{subfig:coq1:ifc}).

Another interesting observation is that even though fuzzers typically
spend more time per generated input, as the underlying types
are more complex and large, mutating the input takes less time
than generating a new one. For IFC, the type-based generator
takes four times longer per input than the type-based fuzzer.

\subsection{Validation and improvement of fuzzers}\label{sec:expfuzz}

The conclusion of~\citet{FuzzChick} seems to hold~--- that is,
FuzzChick shows promise compared to type-based approaches, but has a
long way to go before catching up with the effectiveness of
precondition-driven ones. This led us to wonder, {\em could we further
    improve the performance of FuzzChick using \name?}

We focused on two different aspects of fuzzing: {\em size} and {\em
    feedback}. FuzzChick's generation strategy started small but quickly
increased to quite large sizes, relying on the idea of
``combinatorial advantage'' discussed in
\sectionref{sec:haskell:size}~--- i.e., that larger inputs contain
exponentially many smaller inputs and are therefore more effective for
testing. As we saw there, that is not always the case. After realizing
this, we switched to a more gently increasing size bound which led to
significant improvements in terms of throughput, positively impacting
our bug-finding ability.

With respect to feedback, by using \name{} to evaluate FuzzChick
across multiple workloads we were able to identify, isolate, and fix a
bug that caused it to saturate the seed pool with uninteresting
inputs. FuzzChick (like Zest~\citep{Zest}) keeps two
seed pools: one for valid and one for invalid inputs. FuzzChick's bug
applied to the latter one, and was hidden from its authors as the
variational fuzzer strategy they employed readily gives access to
valid inputs (which are prioritized).

\begin{figure}[htb!]
  \centering
  \begin{subfigure}{\bargraphwidth}
    \centering
    \includegraphics[width=\textwidth]{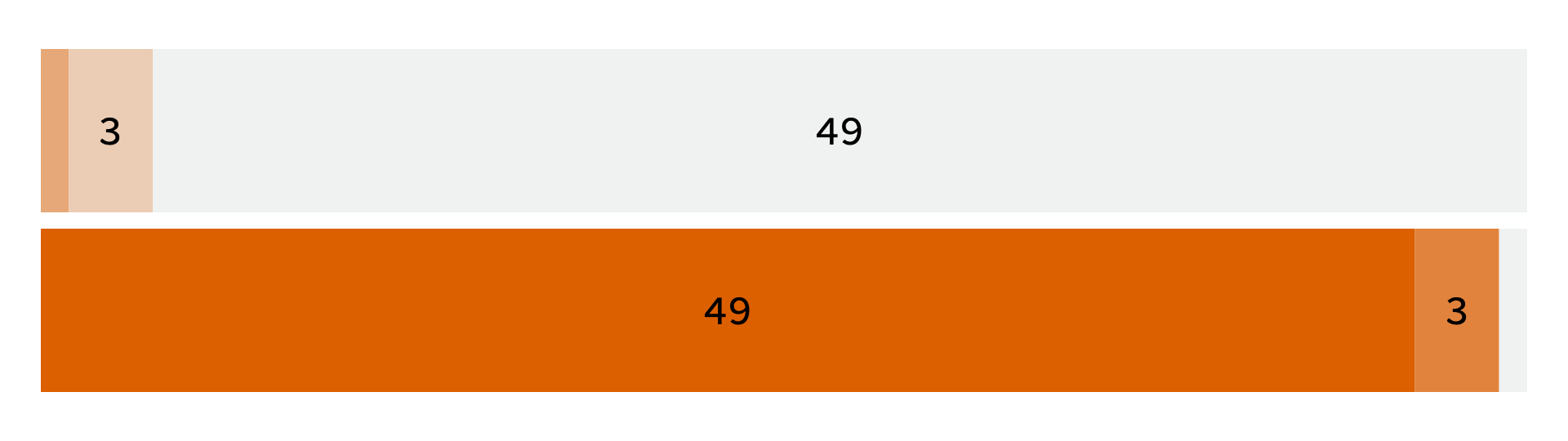}
    \vspace*{-5ex}
    \caption{BST}
  \end{subfigure}
  \begin{subfigure}{\bargraphwidth}
    \centering
    \includegraphics[width=\textwidth]{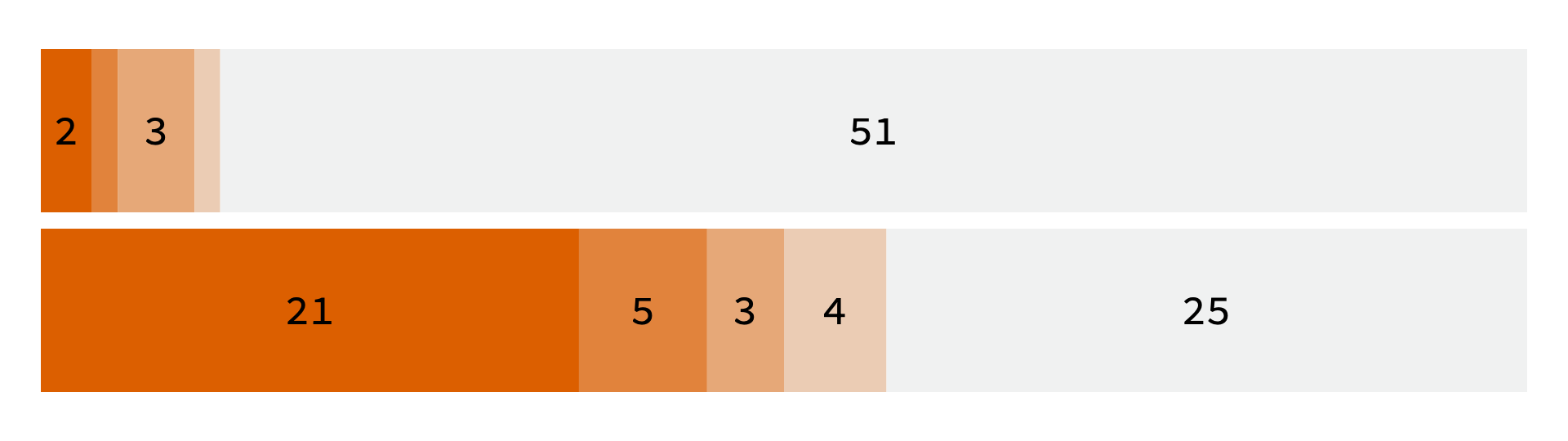}
    \vspace*{-5ex}
    \caption{RBT}
  \end{subfigure}
  \begin{subfigure}{\bargraphwidth}
    \centering
    \includegraphics[width=\textwidth]{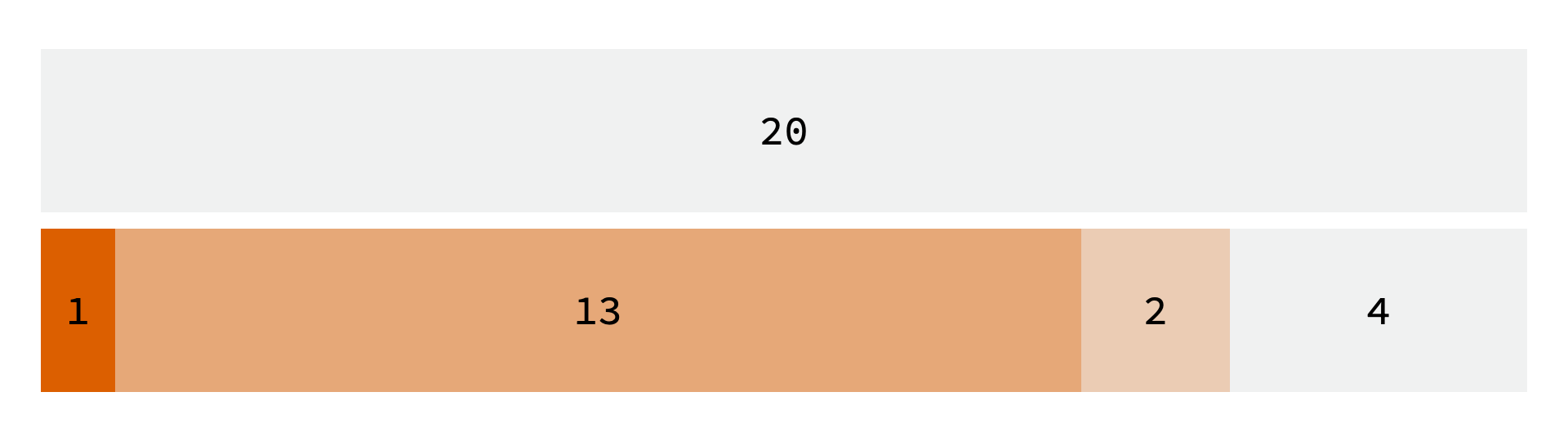}
    \vspace*{-5ex}
    \caption{STLC}
  \end{subfigure}

  \caption{Comparison of the original FuzzChick (top) with the tuned one (bottom).}
  \label{fig:fuzzercomp}
\end{figure}

\ourparagraph{Results}
Figure~\ref{fig:fuzzercomp} demonstrates the bug-finding capabilities
of the original (top) and tuned (bottom) versions of FuzzChick across
the new workloads. The tuned version clearly outperforms the original
in all cases---and is what was used in the previous section.

\ourparagraph{Hardening QuickChick's implementation}
In our experimentation with \name{}, we stress tested some of QuickChick's
features in ways that occasional user interactions could not
hope to reach.
%
One particular bug stood out: The main fuzzing loop of FuzzChick is
written in Gallina and uses a natural number fuel to satisfy the
termination checker. That natural number is extracted as an OCaml
integer for efficiency purposes. However, when extracted, a pattern
match becomes a call to an eliminator:

\smallskip
\noindent
\begin{minipage}{0.33\textwidth}
  \begin{rocqcode}
    Fixpoint loop fuel ... :=
    match fuel with
    | O => (* base *)
    | S fuel' => (* rec *)
    end.
  \end{rocqcode}
\end{minipage}
\quad \quad \quad
\begin{minipage}{0.5\textwidth}
  \begin{ocamlcode}
    let rec loop fuel ... = (fun fO fS n ->
    if n = 0 then fO () else fS (n-1))
    (fun () -> (* base *))
    (fun fuel -> (* rec *))
    fuel
  \end{ocamlcode}
\end{minipage}
\smallskip

\noindent
Can you spot the problem? The extracted version is no longer
identified by the OCaml compiler as tail recursive... which means that
when \name{} used large fuel values, it led to stack overflows!


\section{Experiments: OCaml}\label{sec:ocaml}

Whereas QuickChick is the defacto property-based testing framework in
Rocq, programmers in OCaml have a choice of frameworks: QCheck,
offering a standard QuickCheck-like monadic API for writing
generators; Crowbar, offering fuzzing capabilities as a wrapper around
AFL; and Base\_quickcheck, leveraging the Core standard library
replacement.

For the workloads, we ported the three basic ones from the
previous sections---BST, and RBT, and STLC---to OCaml.
As neither of the three frameworks provides machinery for
automatically deriving either type-based (in the style of Haskell's
generic-random) or specification-based (in the style of Rocq's
QuickChick) generators, we also ported a type-based and a bespoke
generator to each framework.

\renewcommand{\bargraphwidth}{0.45\textwidth}
\begin{figure}[tb]
  \centering
  \begin{subfigure}{\bargraphwidth}
    \centering
    \includegraphics[width=\linewidth]{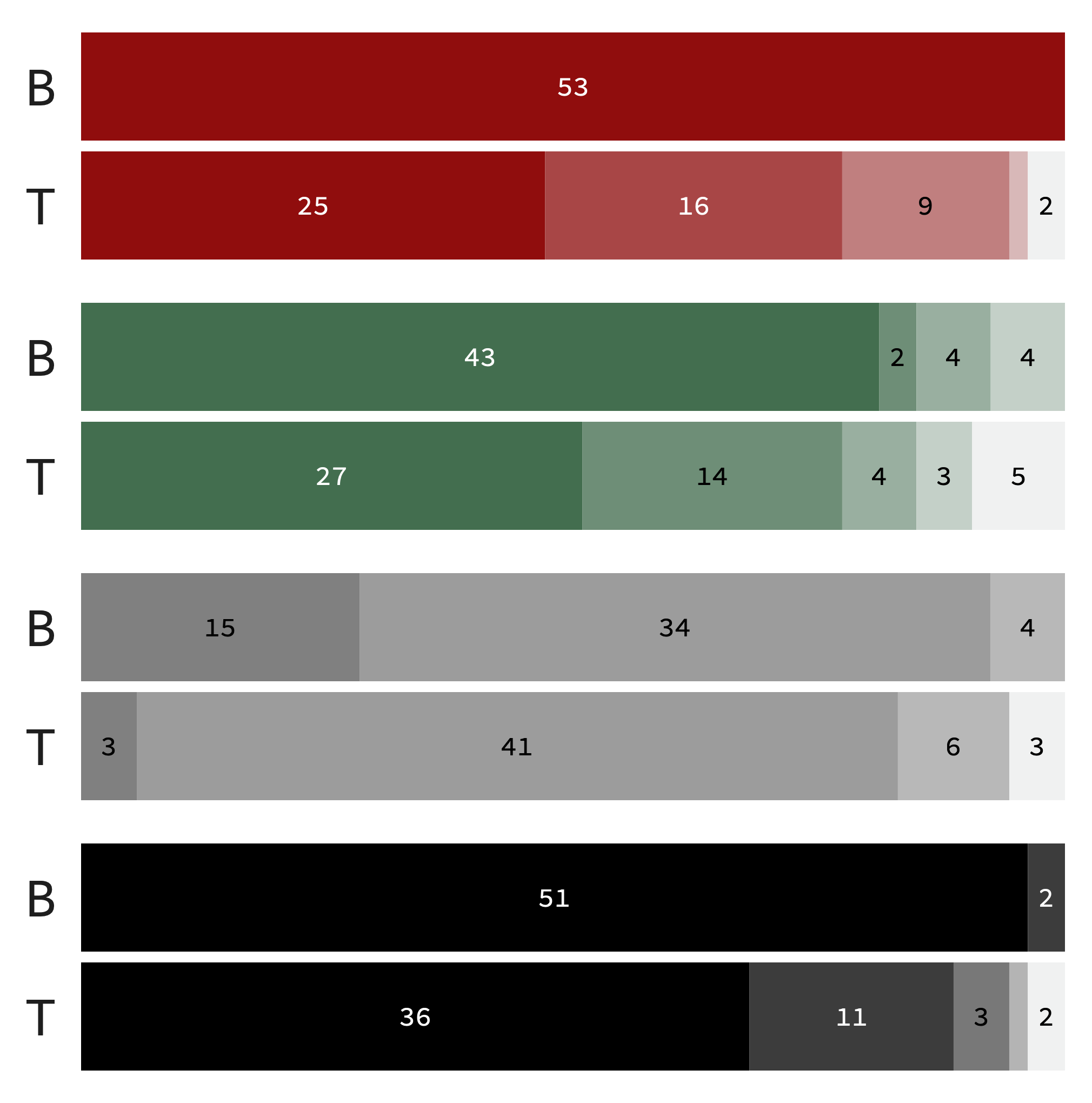}
    \caption{BST}
  \end{subfigure}
  \begin{subfigure}{\bargraphwidth}
    \centering
    \includegraphics[width=\linewidth]{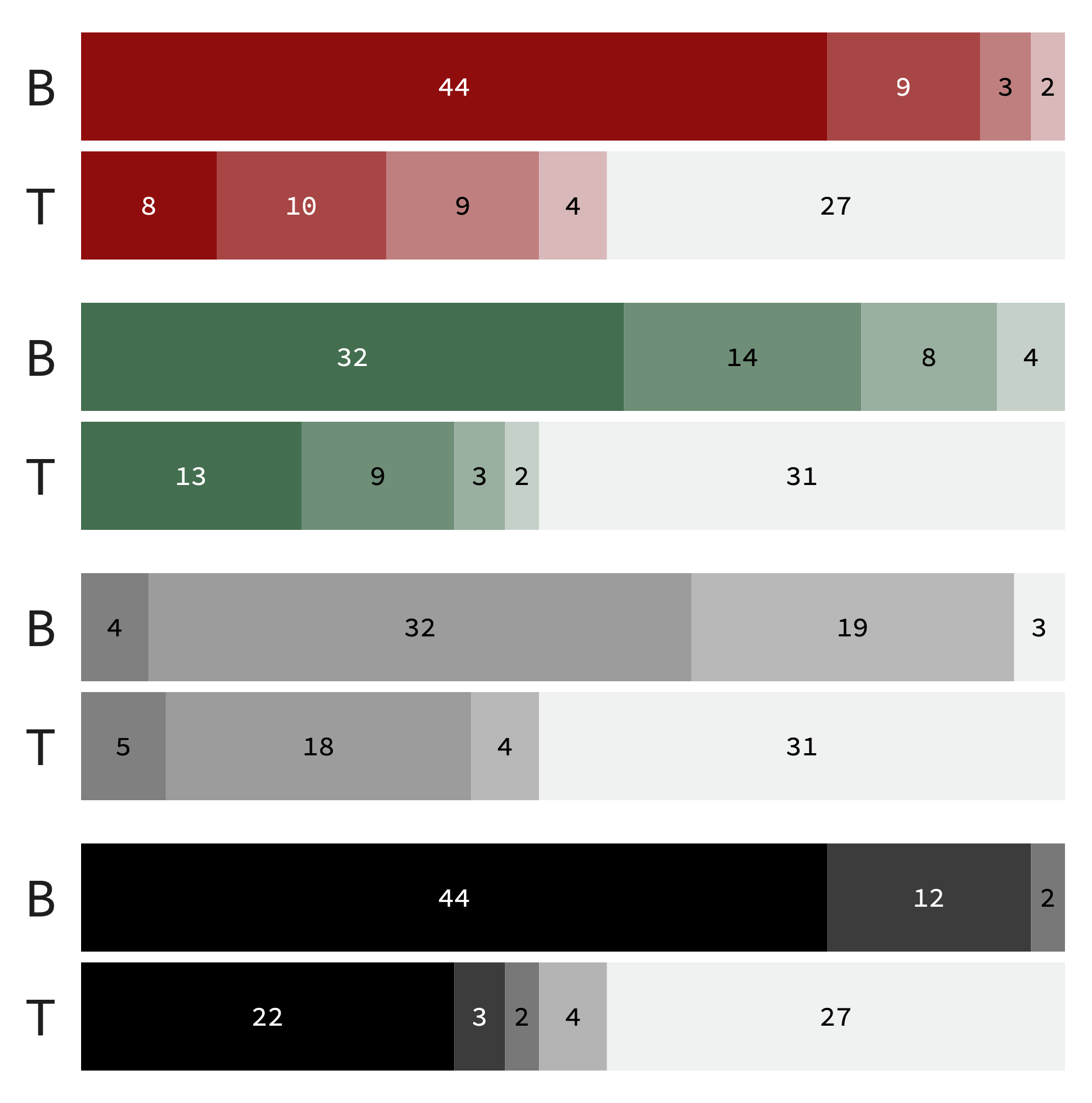}
    \caption{RBT}
  \end{subfigure}

  \begin{subfigure}{\bargraphwidth}
    \centering
    \includegraphics[width=\linewidth]{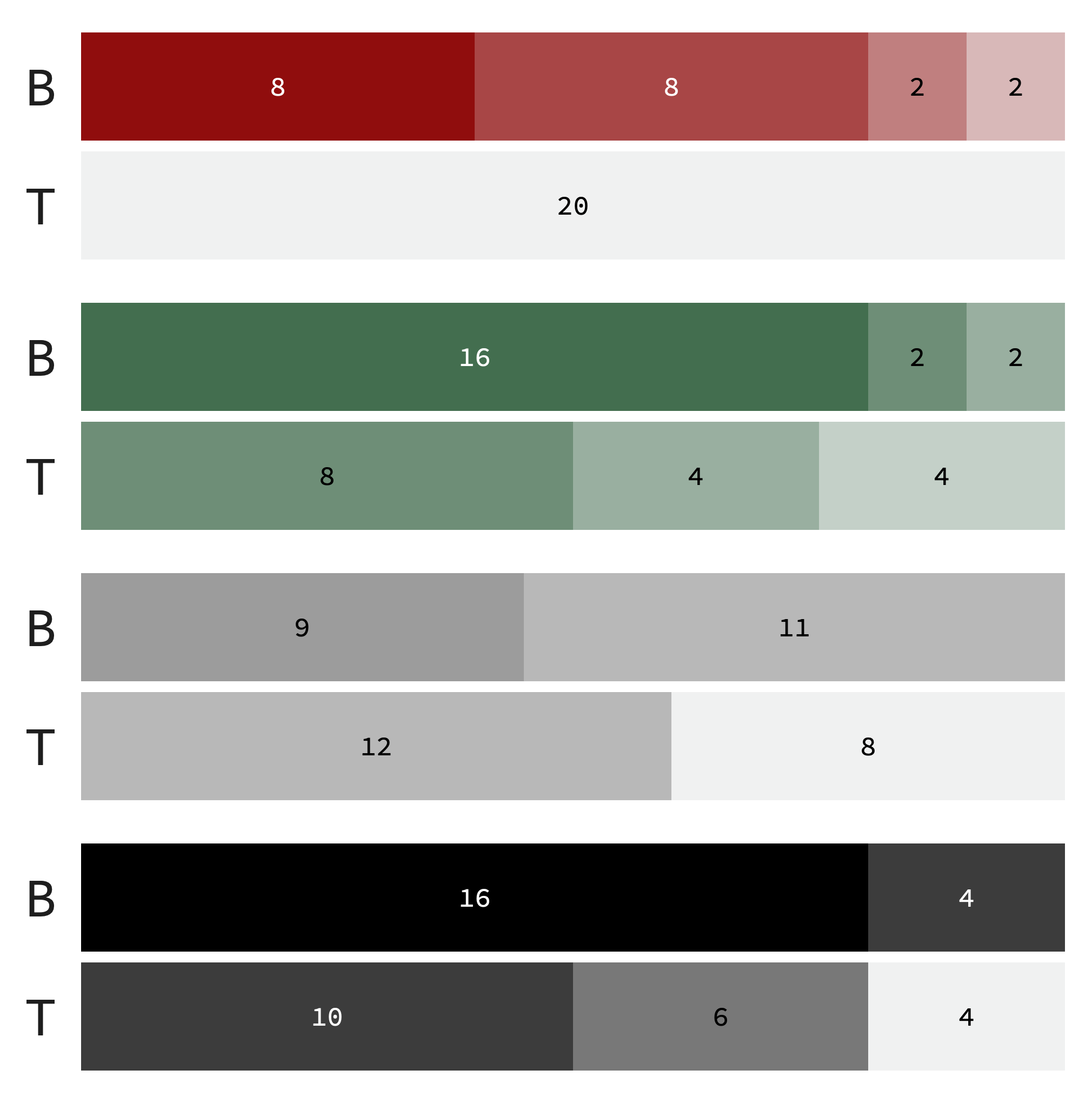}
    \caption{STLC}
  \end{subfigure}
  \caption{
    Effectiveness of OCaml generation strategies on three
    workloads. The first and second buckets for each framework represent the
    bespoke and type-based generators respectively.\\
    \explaincolor{chart_red}{QCheck},
    \explaincolor{chart_green}{Crowbar (random)},
    \explaincolor{gray}{Crowbar (AFL)},
    \explaincolor{chart_black}{Base\_quickcheck}.}
  \label{fig:ocamlcomp}
\end{figure}

\paragraph*{Controlling for size} As our experiments earlier in the paper
showed, the size of generated inputs can have a significant impact on
the effectiveness of testing. However, the three OCaml frameworks
offer vastly diverse default size distributions.
In particular, the one for Base\_quickcheck was similar to the
effective ones from \sectionref{sec:haskell:size}, so we left it as
is.
%
For QCheck, to avoid generating a distribution that is heavily bimodal
(i.e. with many trees containing one or two nodes and most others
containing several thousand) or where the range of integers is too
large (to provoke bugs that rely on collisions), we used the variants
of QCheck's integer generators that focus on smaller integer ranges.
%
Finally, for Crowbar, the default behavior of its \verb|list|
generator (which we use as a basic building block in multiple
strategies) produces very small lists\footnote{Given an input size, it
  will generate an empty list 50\% of the time, or a cons cell whose
  tail is recursively generated with the size parameter halved. In
  practice, that means vanishingly few lists of length more than 5 will
  be generated even for large input sizes.}.
As a result, we re-implemented a list generator using the rest of Crowbar's
API to construct longer lists.


Figure~\ref{fig:ocamlcomp} shows the results in bucket-chart form,
where we used type-based and bespoke generators for each of the frameworks,
using Crowbar both with its purely random and AFL-powered backends.
For these workloads, the Core-library powered Base\_quickcheck bespoke
generators outperform the other frameworks in almost all situations.
The
\href{https://github.com/alpaylan/etna-cli/blob/docs-bst/workloads/OCaml/BST/lib/generators/gen_bespoke_crowbar.ml}{best generator} that we could write using Crowbar's interface performed
the worst out of the strategies we tried. However, that does not mean
that Crowbar as a framework is less effective: rather, just as the
FuzzChick case (\sectionref{sec:rocq:fuzz}), if one takes the effort
to handcraft bespoke generators that satisfy a property's precondition
by construction, coverage-guided fuzzing only adds overhead for
minimal gain.


\section{Experiments: Cross-language comparison of PBT frameworks}
\label{sec:cross}

As we demonstrated throughout this paper, \name{} allows for running
complex experiments that can provide powerful insights to
property-based testing practitioners or framework developers. However,
the scope of the experiments we have provided thus far has been limited
to the level of a single programming language. Given two testing
frameworks or generation strategies within the same language, we can
measure and compare their bugfinding performance across the different
\name{} workloads implemented for that language.

Yet, such intra-language experimentation does not encompass the
current practice of property-based testing, where generation
strategies in one language are used to test systems in another. For
one example, a specification-derived generator for well-typed System F
terms written in Rocq was used to test a higher-order blockchain
language implemented in OCaml~\citep{BlockchainTesting}; for another,
a bespoke generator for file-system interactions written in Erlang
were used to test Dropbox's Python-based file synchronization
service~\citep{MysteriesOfDropbox2016}.
%
If we want \name{} to enable prototyping of and experimentation with
effective testing strategies in practice, we need to be able to
compare the performance of such strategies across languages.
%
%
%

To that end, we developed support to perform cross-language experiments
in \name{}, decoupling generation of inputs and testing of properties.
On the generation side, each aspiring \name{} user must implement their
generation strategy, just like before, but instead of linking it
directly with a framework-dependent way of executing a test, they
need to only output a list of serialized inputs together with the time
it took to generate each one.
On the property end, we have created {\em runners} for the BST, RBT,
and STLC workloads that can read serialized inputs from the command line
to test their related properties with.
\name{} can then perform its analysis in a language-agnostic manner:
it can give precise, fine-grained feedback about the performance of
each generation strategy (without aggregating generation, execution,
or shrinking times together).

\begin{figure}[t]
  \centering
  \begin{tikzpicture}[scale=0.85, every node/.style={scale=0.85},
      genbox/.style={rectangle, draw=chart_blue, fill=chart_blue!10,
          minimum width=1.8cm, minimum height=0.8cm, font=\small\sffamily, rounded corners=2pt, thick, align=center},
      runbox/.style={rectangle, draw=chart_green, fill=chart_green!10,
          minimum width=1.8cm, minimum height=0.8cm, font=\small\sffamily, rounded corners=2pt, thick, align=center},
      serbox/.style={rectangle, draw=chart_orange, fill=chart_orange!10,
          minimum width=1.8cm, minimum height=0.8cm, font=\small\sffamily, rounded corners=2pt, thick, align=center},
      resbox/.style={rectangle, draw=chart_purple, fill=chart_purple!10,
          minimum width=1.8cm, minimum height=0.8cm, font=\small\sffamily, rounded corners=2pt, thick, align=center},
      arrow/.style={->, >=stealth, thick, chart_black},
      datalabel/.style={font=\scriptsize\sffamily},
      timelabel/.style={font=\scriptsize\sffamily, chart_red}
    ]

    \node[font=\sffamily\bfseries] at (-4, 1.2) {Intra-language};
    \node[genbox] (gen1) at (-5.5, -0.25) {Generator};
    \node[runbox] (run1) at (-2.5, -0.25) {Runner};
    \node[resbox] (res1) at (-4, -2) {Result};

    \draw[arrow] (gen1) -- node[above, datalabel] {inputs} (run1);
    \draw[arrow] (run1) -- (res1);

    \draw[decorate, decoration={brace, amplitude=4pt, raise=2pt}, chart_red, thick]
    (-6.3, 0.4) -- (-1.7, 0.4) node[midway, above=6pt, timelabel] {combined time};

    \node[draw=gray!50, dashed, rounded corners=3pt, fit=(gen1)(run1), inner xsep=14pt, inner ysep=6pt] {};

    \node[font=\sffamily\bfseries] at (3.75, 1.2) {Cross-language};
    \node[genbox] (gen2) at (0, -0.25) {Generator};
    \node[serbox] (ser2) at (2.5, -0.25) {Serializer};
    \node[runbox] (read2) at (5, -0.25) {Reader};
    \node[runbox] (run2) at (7.5, -0.25) {Runner};
    \node[resbox] (res2) at (5.5, -2) {Result};

    \draw[arrow] (gen2) -- node[above, datalabel] {inputs} (ser2);
    \draw[arrow] (ser2) -- node[above, datalabel] {S-expr} (read2);
    \draw[arrow] (read2) -- node[above, datalabel] {inputs} (run2);
    \draw[arrow] (run2) -- (res2);

    \draw[decorate, decoration={brace, amplitude=4pt, raise=2pt}, chart_red, thick]
    (-0.9, 0.4) -- (0.9, 0.4) node[midway, above=6pt, timelabel] {gen time};

    \draw[decorate, decoration={brace, amplitude=4pt, raise=2pt}, chart_red, thick]
    (6.7, 0.4) -- (8.3, 0.4) node[midway, above=6pt, timelabel] {exec time};

    \draw[gray!50, dashed, thick] (3.75, -2.3) -- (3.75, 0.8);
    \node[font=\scriptsize\sffamily, gray] at (3, -1) {Lang A};
    \node[font=\scriptsize\sffamily, gray] at (4.5, -1) {Lang B};

  \end{tikzpicture}
  \caption{Intra-language vs cross-language workflows. Left: generator and runner in same language with combined timing. Right: generator serializes inputs; separate reader/runner enables decoupled timing metrics across language boundaries.}
  \label{fig:workflow-comparison}
\end{figure}
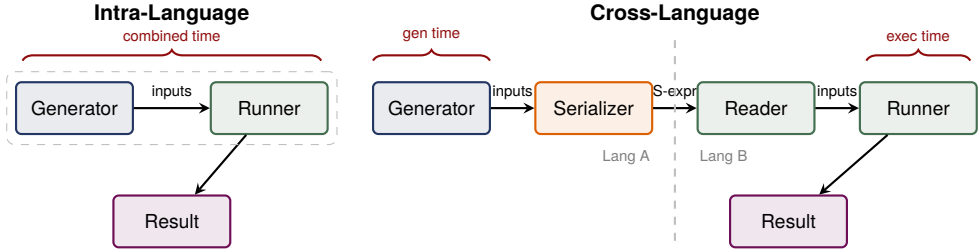

As a beneficial side effect of decoupling strategies from workloads,
the extensibility of \name{} is greatly improved. Integrating a new
language within \name{} no longer {\em requires} porting all of its
workloads in yet another language---although that is still an option.
%
Instead, workloads only need to be implemented once, in any language that
can deserialize inputs to interface with \name{}'s API.
%
And strategies written in a previously unsupported language need only
implement a generator and a serializer to use the existing workloads
for experimentation.
Moreover, the decoupled approach to generation and testing allows for
quick validation of ports of strategies and workloads to new
languages: if running the same generator produces different results in
the cross-language mode than intra-language mode, that points towards
an error in the implementation of the workload or the strategy.
Depictions of the intra-language and cross-language
workflows of \name{} are presented in Fig.~\ref{fig:workflow-comparison}.

\renewcommand{\bargraphwidth}{0.48\textwidth}

\begin{figure}[t]
  \centering
  \begin{subfigure}{\bargraphwidth}
    \centering
    \includegraphics[width=\textwidth]{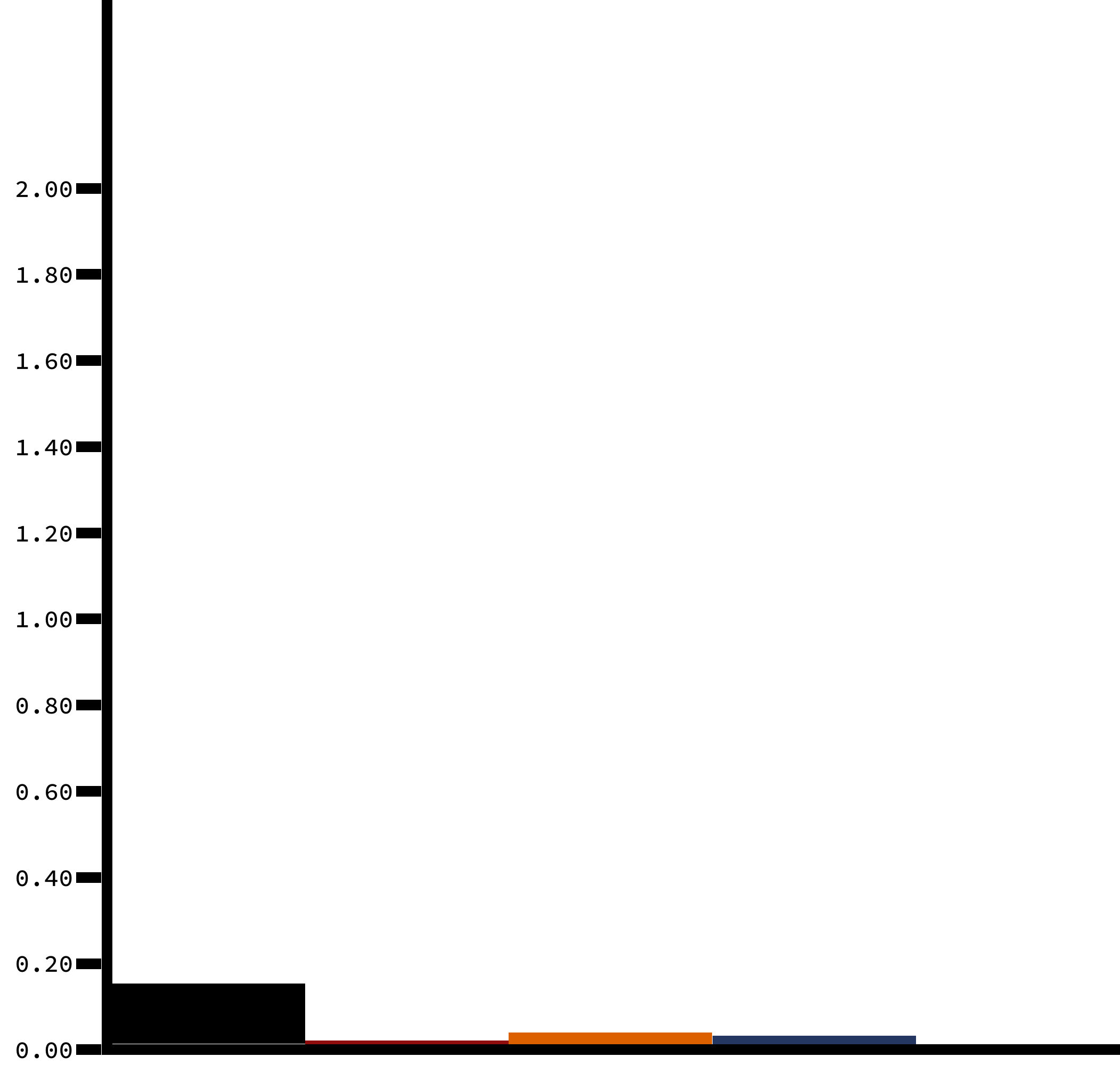}
    \caption{BST}
  \end{subfigure}
  \begin{subfigure}{\bargraphwidth}
    \centering
    \includegraphics[width=\textwidth]{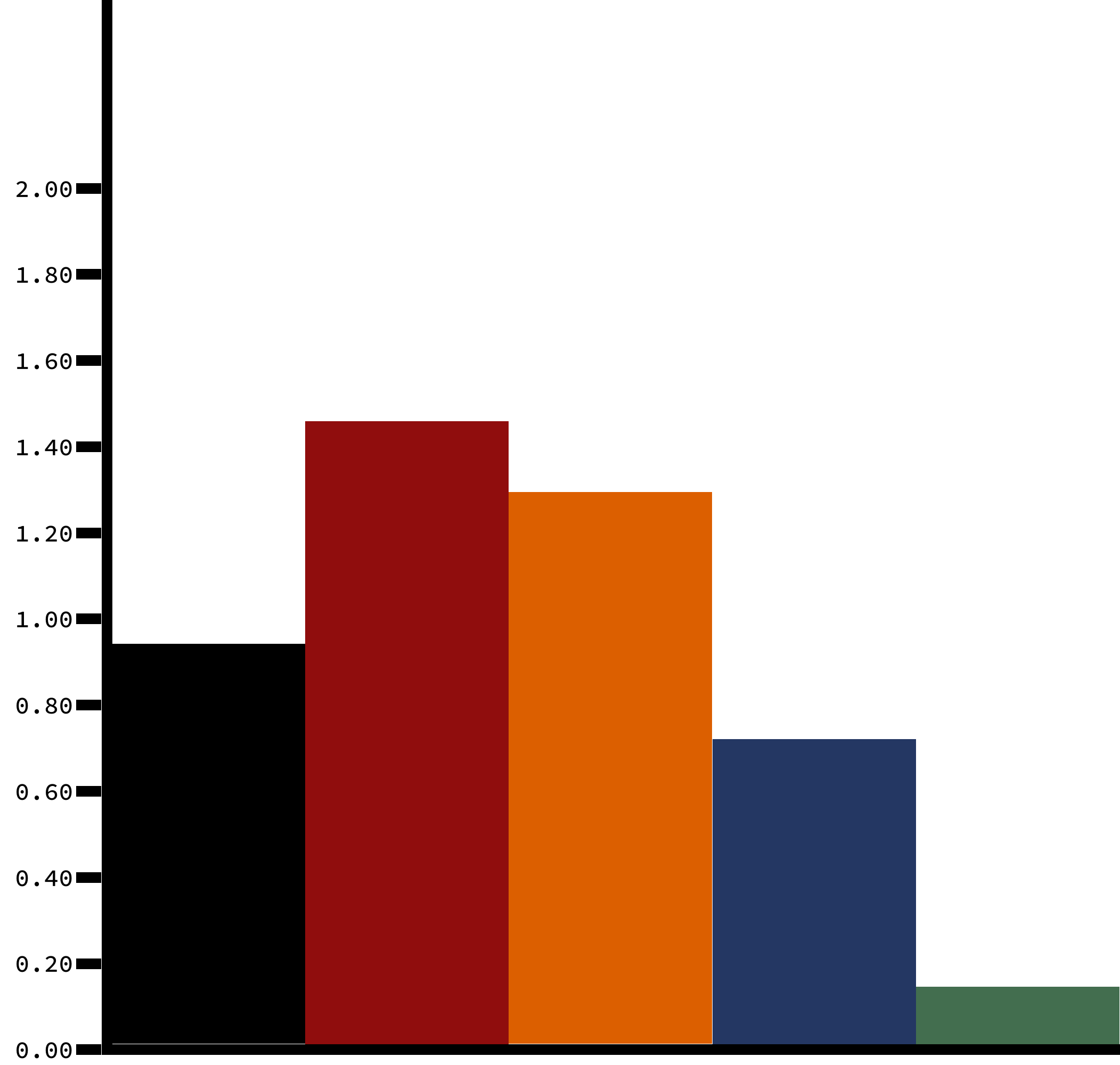}
    \caption{RBT}
  \end{subfigure}

  \begin{subfigure}{\bargraphwidth}
    \centering
    \includegraphics[width=\textwidth]{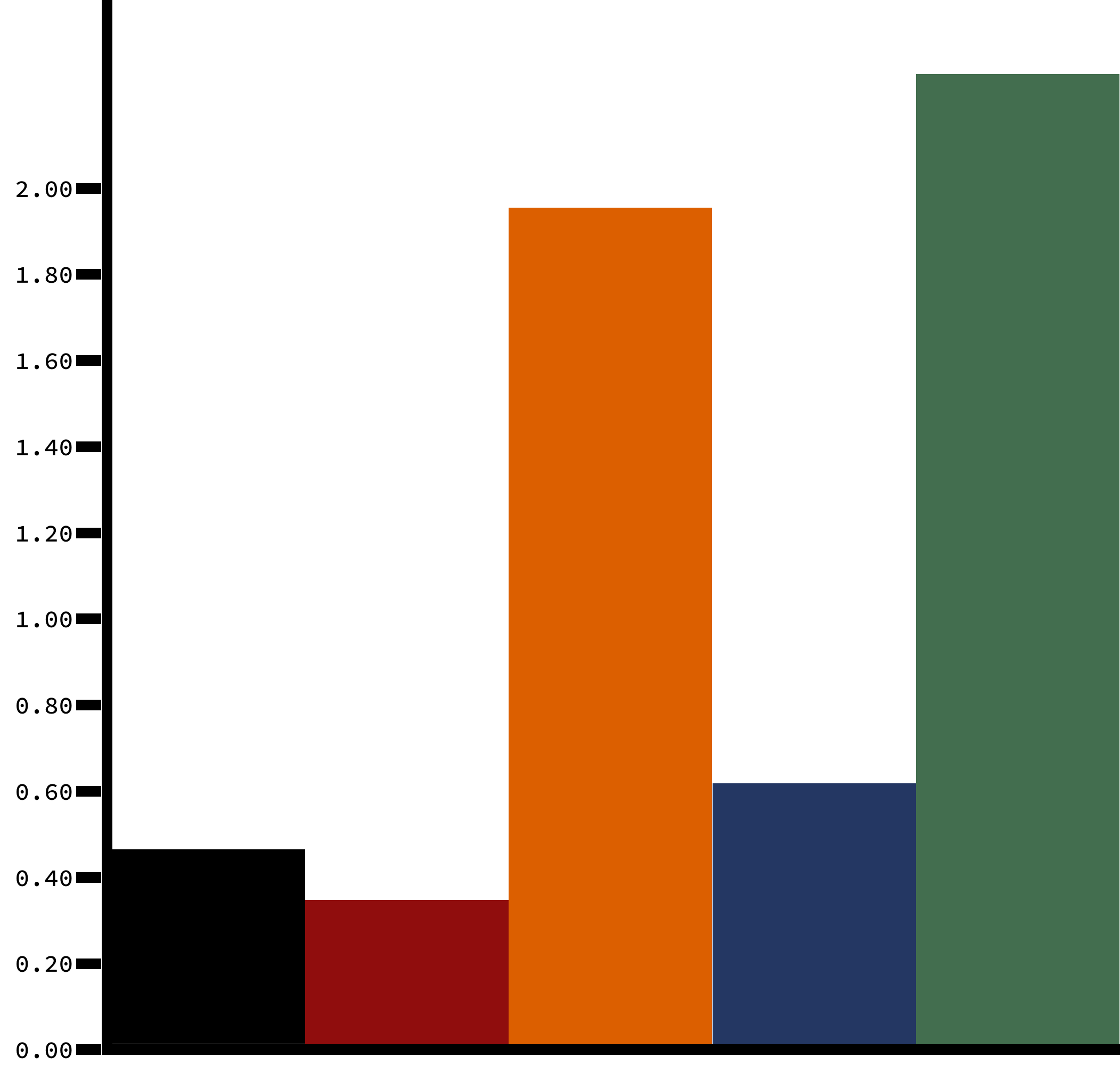}
    \caption{STLC}
  \end{subfigure}


  \captionsetup{justification=centering}
  \caption{
    Generation time to failure (in seconds) for bespoke generators written in different languages.\\
    \explaincolor{chart_black}{Haskell -- QuickCheck},
    \explaincolor{chart_red}{Rocq -- QuickChick},
    \explaincolor{chart_orange}{OCaml -- QCheck},\\
    \explaincolor{chart_blue}{Racket -- Rackcheck},
    \explaincolor{chart_green}{Rust -- QuickCheck}.
  }
  \label{fig:large-cross}
\end{figure}

To demonstrate \name{}'s new capabilities, we pit bespoke generators
written in Haskell, Rocq, OCaml, Racket, and Rust against the three
workloads. The results are shown in Figure~\ref{fig:large-cross}. As
we are dealing with tuned bespoke generators, bucket charts are not
ideal for discerning differences---all generators find basically all
bugs, quickly. 
Instead, the top of the figure shows the {\em total generation time to failure}
per-framework per-workload.

Despite all bespoke generators implementing the same strategy in
principle, there are slight differences in efficiency: the way each
framework manipulates the size of generated inputs across runs and the
performance characteristics of the language both affect the end
result. Still, all strategies are relatively close, with Rust's
QuickCheck, for example, being simultaneously the fastest in the BST
and RBT workloads, and the slowest in STLC (we conjecture because
of excessive creation of expensive closures in binds).

Another takeaway from Figure~\ref{fig:large-cross} is the difficulty of
the workloads themselves: BST is comprised of 52 tasks, and the
slowest generator (Haskell's QuickCheck) still finds all 52 injected
bugs in 130ms; on the other hand, STLC is comprised of only 20 tasks,
but the fastest generator (Rocq's QuickChick) takes 300ms, where the
slowest takes just over 2 seconds.

Moving forward, armed with cross-language evaluation capabilities and
only needing to implement runners once, we hope to rapidly expand the
number of workloads and frameworks available for experimentation.

\section{Related and future work}\label{sec:relwork}
The future directions we imagine for \name{} are inspired by related
work in the literature. Thus, we discuss both related and future work
together in this section.

\name's name, referencing every crossword-puzzler's favorite Italian volcano,
was inspired by two existing benchmark suites in the fuzzing
space: LAVA~\citep{Lava} and Magma~\citep{Hazimeh:2020:Magma}.  Both
provide a suite of workloads that can be used to compare
different fuzzing tools: LAVA's workloads consist of programs with
illegal memory accesses that are automatically injected, while Magma
relies on real bugs forward-ported to the current versions of
libraries. More recently, FixReverter~\citep{FixReverter} offered a
middle ground, generalizing real bug-fixes into patterns and applying
them to multiple locations in a program. \name{} is different from
these suites in a few ways. First, \name{} aims to be a platform for
exploration and evaluation rather than a rigid set of benchmarks.
Thus, we do not claim that \name's workloads are complete~--- instead,
we intend for users to add more over time.  Additionally, evaluating
fuzzing is quite different from evaluating PBT, since PBT is expected
to run for less time on programs with higher input complexity.  This
means that \name's measurement techniques and workload focus must
necessarily be different from LAVA's or Magma's.  Still, there are
ideas worth borrowing from these suites: fuzzing benchmarks generally
record code-coverage information, which we plan
for \name{} to eventually offer as well.

Besides LAVA and Magma, there is a massive literature of Haskell and
Rocq papers from which we will continue to draw both workloads and
frameworks. With the help of the community, we hope \name{} will eventually
include frameworks like: Luck~\citep{Luck}, a
language for preconditions from which generators can be inferred;
FEAT~\citep{Duregard12}, an enumerator framework focusing on
uniformity; tools for deriving better Haskell
generators~\citep{MistaR19A,MistaR19B}; and specification-driven
enumerators for QuickChick~\citep{ComputingCorrectly}.

Outside of the currently supported languages and frameworks, there are yet more opportunities for
growth. We will solicit framework maintainers
and researchers to add support for other languages such as Scala (SciFe~\citep{KurajK14,KurajKJ15} and
ScalaCheck~\citep{ScalaCheck}), Erlang (QuviQ~\citep{Arts2008} or
PropEr~\citep{PapadakisS11}), or Isabelle~\citep{Bulwahn12, Bulwahn12smartgen}.

Finally, the presentation back end of \name{} is fit-for-purpose, but
we intend to do further research into the best possible ways to
visualize PBT results.  Consulting experts in human-computer
interaction, we plan to use tools like
Voyager~\citep{wongsuphasawat_voyager_2017} to explore which kinds of
outcome visualizations real users of
\name{} want. At the very least, integrating \name{} into a Jupyter
notebook~\citep{noauthor_project_2023} and providing hooks into a powerful graphics engine like
Vega-lite~\citep{satyanarayan_vega-lite_2017} would make it easier for users to experiment with
visualizations.

\section{Conclusion}\label{sec:future}
We designed \name{} to meet a concrete need in our research~--- we needed a clear way
to convince ourselves and others that the PBT tools we build are worth pursuing.
\name{} provides that, with an extensible suite of interesting workloads and the
infrastructure necessary to validate and refine designs against them. In
\sectionref{sec:haskell} - \sectionref{sec:cross}, we originally set out to
answer straightforward questions about whether X is better than Y, and while we
did get feedback about general trends, we also uncovered some
unexpected nuances of the testing process. PBT-curious readers may have further
questions building upon and extending beyond our explorations.
\name{} is there for you!

\section*{Acknowledgments}

This work was supported by the NSF under award
\#1955610 \emph{Bringing Python Up to Speed} and under \#2145649
\emph{CAREER: Fuzzing Formal Specifications}.

\bibliographystyle{ACM-Reference-Format}
\bibliography{leo,quick-chick}

\clearpage
\pagebreak
\appendix

\end{document}